\documentclass[english,showpacs,english,aps,prc,twocolumn,superscriptaddress,floatfix]{revtex4}
\usepackage[T1]{fontenc}
\usepackage[latin9]{inputenc}
\usepackage{refstyle}
\usepackage{textcomp}
\usepackage{amstext}
\usepackage{amssymb}
\usepackage{graphicx}
\usepackage{color}
\usepackage{lipsum}
\usepackage{float} 
\usepackage{tabularx}
\usepackage{multirow}
\usepackage{mathtools}
\usepackage{ulem}		
\usepackage{orcidlink} 
\usepackage{enumitem}

\makeatletter

\newcommand{\lyxmathsym}[1]{\ifmmode\begingroup\def\b@ld{bold}
  \text{\ifx\math@version\b@ld\bfseries\fi#1}\endgroup\else#1\fi}

\RS@ifundefined{subref}
  {\def\RSsubtxt{section~}\newref{sub}{name = \RSsubtxt}}
  {}
\RS@ifundefined{thmref}
  {\def\RSthmtxt{theorem~}\newref{thm}{name = \RSthmtxt}}
  {}
\RS@ifundefined{lemref}
  {\def\RSlemtxt{lemma~}\newref{lem}{name = \RSlemtxt}}
  {}

\@ifundefined{textcolor}{}
{%
 \definecolor{BLACK}{gray}{0}
 \definecolor{WHITE}{gray}{1}
 \definecolor{RED}{rgb}{1,0,0}
 \definecolor{GREEN}{rgb}{0,1,0}
 \definecolor{BLUE}{rgb}{0,0,1}
 \definecolor{CYAN}{cmyk}{1,0,0,0}
 \definecolor{MAGENTA}{cmyk}{0,1,0,0}
 \definecolor{YELLOW}{cmyk}{0,0,1,0}
}

\makeatother

\usepackage{babel}
\begin{document}

\title{Isospin diffusion from $^{40,48}$Ca$+^{40,48}$Ca experimental data at Fermi energies: Direct comparisons with transport model calculations}

\author{Q.~Fable \orcidlink{0000-0001-6498-0552}}
\email{quentin.fable@l2it.in2p3.fr}
\affiliation{Laboratoire des 2 Infinis - Toulouse (L2IT-IN2P3), Universit\'e de Toulouse, CNRS, UPS, F-31062 Toulouse Cedex 9, France}

\author{L.~Baldesi}
\affiliation{Dipartimento di Fisica e Astronomia, Universit\`{a} di Firenze, I-50019 Sesto Fiorentino, Italy}
\affiliation{INFN Sezione di Firenze, I-50019 Sesto Fiorentino, Italy}

\author{S.~Barlini}
\affiliation{Dipartimento di Fisica e Astronomia, Universit\`{a} di Firenze, I-50019 Sesto Fiorentino, Italy}
\affiliation{INFN Sezione di Firenze, I-50019 Sesto Fiorentino, Italy}

\author{E.~Bonnet}
\affiliation{SUBATECH UMR 6457, IMT Atlantique, Universit\'e de Nantes, CNRS-IN2P3, 44300 Nantes, France}

\author{B.~Borderie}
\affiliation{Universit\'e Paris-Saclay, CNRS/IN2P3, IJCLab, 91405 Orsay, France}

\author{R.~Bougault}
\affiliation{Normandie Univ, ENSICAEN, UNICAEN, CNRS/IN2P3, LPC Caen, F-14000 Caen, France}

\author{A.~Camaiani}
\affiliation{Dipartimento di Fisica e Astronomia, Universit\`{a} di Firenze, I-50019 Sesto Fiorentino, Italy}
\affiliation{INFN Sezione di Firenze, I-50019 Sesto Fiorentino, Italy}

\author{G.~Casini}
\affiliation{INFN Sezione di Firenze, I-50019 Sesto Fiorentino, Italy}

\author{A.~Chbihi}
\affiliation{GANIL, CEA/DRF-CNRS/IN2P3, Bvd. Henri Becquerel, F-14076 Caen CEDEX, France}

\author{C.~Ciampi}
\affiliation{GANIL, CEA/DRF-CNRS/IN2P3, Bvd. Henri Becquerel, F-14076 Caen CEDEX, France}

\author{J.~A.~Due\~{n}as}
\affiliation{Departamento de Ingenier\'{i}a El\'ectrica y Centro de Estudios Avanzados en F\'{i}sica, Matem\'{a}ticas y Computaci\'{o}n, Universidad de Huelva, 21007 Huelva, Spain} 

\author{J.D.~Frankland}
\affiliation{GANIL, CEA/DRF-CNRS/IN2P3, Bvd. Henri Becquerel, F-14076 Caen CEDEX, France}

\author{T.~G\'enard}
\affiliation{GANIL, CEA/DRF-CNRS/IN2P3, Bvd. Henri Becquerel, F-14076 Caen CEDEX, France}

\author{D.~Gruyer}
\affiliation{Normandie Univ, ENSICAEN, UNICAEN, CNRS/IN2P3, LPC Caen, F-14000 Caen, France}

\author{M.~Henri}
\affiliation{GANIL, CEA/DRF-CNRS/IN2P3, Bvd. Henri Becquerel, F-14076 Caen CEDEX, France}

\author{B.~Hong}
\affiliation{Center for Extreme Nuclear Matters (CENuM), Korea University, Seoul 02841, Republic of Korea}
\affiliation{Department of Physics, Korea University, Seoul 02841, Republic of Korea}

\author{S.~Kim}
\affiliation{Center for Exotic Nuclear Studies, Institute for Basic Science, 55 Expo-ro, Yuseong-gu, Daejeon 34126, Republic of Korea}

\author{A.~J.~Kordyasz}
\affiliation{Heavy Ion Laboratory, Warsaw University, 02-093 Warsaw, Poland}

\author{T.~Kozik}
\affiliation{Institute of Physics, Jagiellonian University, 30-348 Krakow, Poland}

\author{A.~Le~F\`evre}
\affiliation{GSI Helmholtzzentrum f\"{u}r Schwerionenforschung GmbH, D-64291 Darmstadt, Germany}

\author{N.~Le~Neindre}
\affiliation{Normandie Univ, ENSICAEN, UNICAEN, CNRS/IN2P3, LPC Caen, F-14000 Caen, France}

\author{I.~Lombardo}
\affiliation{Dipartimento di Fisica e Astronomia, University of Catania, Via Santa Sofia 64, I-95123 Catania, Italy}
\affiliation{INFN Sezione di Catania, Via Santa Sofia 64, I-95123, Catania, Italy}

\author{O.~Lopez}
\affiliation{Normandie Univ, ENSICAEN, UNICAEN, CNRS/IN2P3, LPC Caen, F-14000 Caen, France}

\author{T.~Marchi}
\affiliation{INFN Laboratori Nazionali di Legnaro, Viale Dell'Universit\`{a} 2, 35020 Legnaro (PD) Italy}

\author{P.~Marini}
\affiliation{Univ. Bordeaux, CNRS, LP2I, UMR 5797, F-33170 Gradignan, France}
\affiliation{GANIL, CEA/DRF-CNRS/IN2P3, Bvd. Henri Becquerel, F-14076 Caen CEDEX, France}

\author{S.~H.~Nam}
\affiliation{Center for Extreme Nuclear Matters (CENuM), Korea University, Seoul 02841, Republic of Korea}
\affiliation{Department of Physics, Korea University, Seoul 02841, Republic of Korea}

\author{A.~Ono}
\affiliation{Department of Physics, Tohoku University, Sendai 980-8578, Japan}

\author{J.~Park}
\affiliation{Center for Extreme Nuclear Matters (CENuM), Korea University, Seoul 02841, Republic of Korea}
\affiliation{Department of Physics, Korea University, Seoul 02841, Republic of Korea}

\author{M.~P\^arlog}
\affiliation{Normandie Univ, ENSICAEN, UNICAEN, CNRS/IN2P3, LPC Caen, F-14000 Caen, France}
\affiliation{National Institute for Physics and Nuclear Engineering, RO-077125 Bucharest-M\u{a}gurele, Romania}

\author{S.~Piantelli}
\affiliation{INFN Sezione di Firenze, I-50019 Sesto Fiorentino, Italy}

\author{A.~Rebillard-Souli\'e}
\affiliation{Normandie Univ, ENSICAEN, UNICAEN, CNRS/IN2P3, LPC Caen, F-14000 Caen, France}

\author{G.~Verde}
\affiliation{Laboratoire des 2 Infinis - Toulouse (L2IT-IN2P3), Universit\'e de Toulouse, CNRS, UPS, F-31062 Toulouse Cedex 9, France}
\affiliation{INFN Sezione di Catania, Via Santa Sofia 64, I-95123, Catania, Italy}

\author{E.~Vient}
\affiliation{Normandie Univ, ENSICAEN, UNICAEN, CNRS/IN2P3, LPC Caen, F-14000 Caen, France}

\collaboration{INDRA and INDRA-FAZIA collaborations}\noaffiliation

\begin{abstract}

This article presents an investigation of isospin equilibration in cross-bombarding $^{40,48}$Ca$+^{40,48}$Ca reactions at $35$ MeV/nucleon, by comparing experimental data with filtered transport model calculations. 
Isospin diffusion is studied using the evolution of the isospin transport ratio with centrality. 
The asymmetry parameter $\delta=(N-Z)/A$ of the quasiprojectile (QP) residue is used as isospin-sensitive observable, while a recent method for impact parameter reconstruction is used for centrality sorting.
A benchmark of global observables is proposed to assess the relevance of the antisymmetrized molecular dynamics (\textsc{amd}) model, coupled to \textsc{gemini++}, in the study of dissipative collisions. Our results demonstrate the importance of considering cluster formation to reproduce observables used for isospin transport and centrality studies.
Within the \textsc{amd} model, we prove the applicability of the impact parameter reconstruction method, enabling a direct comparison to the experimental data for the investigation of isospin diffusion.
For both, we evidence a tendency to isospin equilibration with an impact parameter decreasing from $9$ to $3$ fm, while the full equilibration is not reached.
A weak sensitivity to the stiffness of the equation of state employed in the model is also observed, with a better reproduction of the experimental trend for the neutron-rich reactions. 

\end{abstract}
\date{\today}
\maketitle

\section{Introduction}\label{sec_intro}

The nuclear equation of state (EoS) is a major issue in modern astrophysics and nuclear physics, as it is a central ingredient in the modeling of core-collapse supernovae, neutron stars and compact binary stars mergers \cite{LATTIMER2016127}, and also in heavy ion collisions (HICs) dynamics and nuclear structure \cite{Di_Toro_2010}.
Many efforts are nowadays dedicated to the study of the density dependence of the nuclear matter symmetry energy \cite{PhysRevC_86_015803, Lattimer_2013, Huth2022, LYNCH2022137098, Tsang2024}.
This term represents the energy cost of converting all protons in symmetric matter into neutrons (at fixed temperature and density), and it is usually defined as the second derivative of the energy per nucleon in nuclear matter with respect to the neutron-to-proton asymmetry (isospin).
Better knowledge of the symmetry energy is necessary in the context of astrophysical simulations which aim to describe neutron-rich systems over a wide range of densities.
At suprasaturation densities, constraints can be extracted by relativistic HICs \cite{RUSSOTTO2011471} and multimessenger measurements \cite{Abbott_PhysRevLett_119_161101}, interest in which has been renewed by the recent breakthrough in gravitational wave measurements.
At subsaturation, HICs offer a unique opportunity to study the modification of the chemical composition induced by the formation and dissolution of clusters in dilute nuclear matter \cite{Typel_2013}.
The study of these reactions allows one to probe the thermodynamical properties of the expanding nuclear system, thus the effect of cluster formation on the symmetry energy.  
Various HIC observables have been used to study the symmetry energy density dependence of neutron-rich matter.
Experimental probes such as nuclear masses \cite{PhysRevLett_111_232502}, isobaric analog states \cite{DANIELEWICZ20141}, collective flows \cite{PhysRevC_94_034608}, pion yield ratios \cite{PhysRevLett_126_162701} and isospin transport \cite{PhysRevLett_102_122701} have been used to constrain the symmetry energy functional.
More recently, it has been shown that such measurements are not only consistent with astrophysical observations but they also provide important constraints on the nuclear EoS at suprasaturation densities \cite{Huth2022, LYNCH2022137098}.
It was furthermore pointed out that improvements in the interpretation of experimental observables are needed if we want to refine our knowledge of the EoS.  
In this context, isospin transport is particularly interesting as it led to one of the first experimental constraints on the symmetry energy \cite{PhysRevLett_102_122701}.
This process, expected to occur in binary dissipative reactions, corresponds to a stochastic and differential exchange of nucleons between projectile and target.
In particular, in collisions with different projectile and target $N/Z$ asymmetries, a balancing in the neutron richness of different isospin regions is expected to take place, leading to a rearrangement of the neutron-to-proton ratio of the colliding nuclei, called isospin diffusion.
The degree of isospin equilibration, directly related to the magnitude of the symmetry energy (at a given local density), can be experimentally estimated from the isospin transport ratio \cite{Rami_PhysRevLett_84_1120}, widely used over the years \cite{PhysRevLett_102_122701, Rami_PhysRevLett_84_1120, CAMAIANI_PRC_102_044607, CAMAIANI_PhysRevC_103_014605, PhysRevC_103_014603, Ciampi_PRC_106_024603, Fable_PRC_107_014604}.

Constraints on the EoS are obtained by comparing nuclear experimental data to transport model calculation, where it is possible to test the interplay between the mean-field effects and nucleon-nucleon collisions.
In addition to the inherent difficulty of implementing various algorithms for solving the transport equations, uncertainties also arise from the comparison protocols between the experimental data and the transport models.
Indeed, as most of the relevant observables are not directly measurable experimentally, surrogate variables are usually used, complicating the interpretations.
Suitable comparisons should therefore focus on an identical set of observables, with calculations ideally run over the same domain of impact parameter probed by the experiment and taking into account the intrinsic limits of the experimental setup. 
%
To this aim, this work presents an investigation of isospin diffusion in $^{40,48}$Ca$+^{40,48}$Ca reactions at $35$ MeV/nucleon, from both experimental and transport model simulation data.
The experimental setup is described in Sec. \ref{sec_exp}. 
The model framework is discussed in Sec. \ref{sec_simu} along with global comparisons with the experimental data.
The impact parameter estimation method is presented in Sec. \ref{sec_ip_recon}, while the results on isospin diffusion are presented in Sec. \ref{sec_isodiff}.
A summary is given and conclusions are finally drawn in Sec. \ref{sec_conclusion}.

\section{Experimental setup and event selection}\label{sec_exp}

\subsection{Experimental setup}\label{subsec_setup}
The experiment was performed at the GANIL facility, where beams of $^{40,48}$Ca at 35 MeV/nucleon impinged on self-supporting $1.0$ mg/cm$^{2}$ $^{40}$Ca or $1.5$ mg/cm$^{2}$ $^{48}$Ca targets placed inside the INDRA vacuum chamber, for a typical beam intensity around $5.10^{7}$ pps. 
The detection apparatus consisted of the coupling of the $4\pi$ charged particle array INDRA and the VAMOS high-acceptance spectrometer. 
INDRA is composed of detection telescopes arranged in rings and centered around the beam axis. 
A detailed description of the INDRA detector and its electronics can be found in \cite{I3-Pou95, I5-Pou96}.
For this experiment, INDRA covered polar angles from $7^{\circ}$ to $176^{\circ}$, as rings 1 to 3 were removed to allow the mechanical coupling with VAMOS in the forward direction. 
Rings 4 to 9 ($7^{\circ}-45^{\circ}$) consisted each of 24 three-layer detection telescopes: a gas-ionization chamber operated with C$_{3}$F$_{8}$ gas at low pressure, a 300 or 150 $\mu$m silicon wafer, and a CsI(Tl) scintillator (14 to 10 cm thick) read by a photomultiplier tube. 
Rings 10 to 17 ($45^{\circ}-176^{\circ}$) included 24, 16, or 8 two-layer telescopes: a gas-ionization chamber and a CsI(Tl) scintillator of 8, 6, or 5 cm thickness. 
Fragment identification thresholds were about $0.5$ and $1.5$ MeV/nucleon for the lightest ($Z \lesssim 10$) and the heaviest fragments, respectively. 
INDRA allowed charge and isotope identification up to boron and charge identification for heavier fragments. 

VAMOS is composed of two large magnetic quadrupoles focusing the incoming ions in the vertical and horizontal planes, followed by a large magnetic dipole \cite{PULLANHIOTAN2008343, LEMASSON2023168407}.
In the used configuration, the spectrometer was rotated to $4.5^{\circ}$ with respect to the beam axis, to cover the forward polar angles from $2.56^{\circ}$ to $6.50^{\circ}$, favoring the detection of a fragment emitted slightly above the grazing angles of the studied reactions. 
The momentum acceptance was about $\pm5\%$ and the focal plane was located $9$ m downstream the target, giving a large enough time of flight (ToF) base to obtain a mass resolution of about $\Delta A/A \approx 1/165$ for the isotopes produced in the collisions. 
The VAMOS detection setup included two position-sensitive drift chambers used to determine the trajectories of the reaction products at the focal plane, followed by a seven-module ionization chamber, a 500 $\mu$m thick Si wall (18 independent modules), and a $1$ cm thick CsI(Tl) wall (80 independent modules), allowing the measurements of the ToF, energy loss ($\Delta E$), and energy ($E$) parameters. 
Around 12 magnetic rigidity settings, from $0.661$ to $2.220$ Tm, were used for each system to cover the full velocity range of the fragments.
At least one hit on the VAMOS silicon wall was required for each event to be acquired.

An overview of the acquired data was presented in \cite{Fable_PRC_106_024605}, with a detailed description of the identification, reconstruction and event normalization procedures.
In a second paper, a study dedicated to isospin diffusion and migration was presented, demonstrating the potential of the INDRA-VAMOS coupling to provide further experimental constraints on the nuclear symmetry energy \cite{Fable_PRC_107_014604}.

\subsection{Data selection}\label{subsec_datasel}

The following preliminary selections have been applied for the present analysis :

\begin{enumerate}[label=(\roman*)]
	\item Only multiplicity ``$1"$ events in the VAMOS Si wall were considered. This offline selection was applied to make sure that the positions measured in the drift chambers are correct and to remove events with ambiguous trajectory reconstruction. 

	\item Elastic-like events, defined as events with no hit in INDRA and a fragment identical to the projectile in VAMOS, were also removed offline.

	\item The most incomplete events were discarded by requesting a measured total charge $Z_{tot} \geq 10$ and a total parallel momentum (along the beam axis) $p_{z}^{tot} \geq 0.5 \cdot p_{beam}$, where $p_{beam}$ is the beam initial momentum.

	\item Events with a potential quasiprojectile (QP) measured in INDRA were discarded by requiring that $Z_{PLF} > Z_{I}^{max,fwd}$, where $Z_{PLF}$ and $Z_{I}^{max,fwd}$ are respectively the charge of the fragment detected in VAMOS and of the heaviest fragment forward-emitted in the reaction center of mass (c.m.) identified in charge with INDRA. Since we focus on the study of isospin-sensitive observables from the QP remnant, such events are indeed not relevant due to the limited mass identification ($Z \lesssim 5$) with INDRA.
\end{enumerate}

In summary, the applied data selections criteria represent a subset of about $90\%$ and $87\%$ of the total statistics of the experimental data for $^{40}$Ca and $^{48}$Ca projectile reactions, respectively.

\section{Simulation codes and global comparisons}\label{sec_simu}

\subsection{Antisymmetrized molecular dynamics}\label{subsec_amd}

The present analysis relies on the comparison of experimental data with the output of the antisymmetrized molecular dynamics (\textsc{amd}) model \cite{ONO_PTP_87_5_1185}.
\textsc{amd} is a microscopic transport model widely used to describe various features of ground state nuclei \cite{KANADAENYO2003497} and out of equilibrium many-body dynamics \cite{Ono_2013}, recently reviewed in \cite{WOLTER2022103962}.

A general issue encountered when comparing transport model simulations with experimental HIC data at Fermi energies is the underestimation of light cluster yields. 
For example, in a recent study Frosin \textit{et~al.} investigated the role of clustering at Fermi energies by comparing $^{32}$S$+^{12}$C and $^{20}$Ne$+^{12}$C reactions at 25 and 50 MeV/nucleon measured with four blocks of the FAZIA detector array with the output of the \textsc{amd} model \cite{PhysRevC_107_044614}.
It was shown that accounting for cluster formation helps to better reproduce experimental multiplicities of both light-charged particles (LCP, $Z\leq 2$) and intermediate mass fragments (IMF, $ 3 \leq Z \leq 8$) and also their kinetic energy distributions.  
Furthermore, to our knowledge, in the literature very few works have addressed the specific issue of clustering in semiperipheral to peripheral collisions, where binary dissipative collisions exhaust a large part of the total reaction cross section \cite{PhysRevC_96_034622}.

In this perspective, we present in this section a comparison of the INDRA-VAMOS data with a basic version of \textsc{amd}, namely \textsc{amd-nc}, and a more recent version with cluster correlations, namely \textsc{amd}-CC.
More precisely, in \textsc{amd}-CC cluster correlations are explicitly taken into account by allowing each of the scattered nucleons to form light clusters ($A\leq4$) and also heavier nuclei (up to $A \lesssim 10$) from the introduction of intercluster correlations as a stochastic process of intercluster binding, as described in \cite{ONO2019139}.

Concerning \textsc{amd-nc}, the same version as the one detailed in \cite{PhysRevC_59_853} was employed, proven to be suitable for the study of Ca+Ca collisions at 35 MeV/nucleon \cite{WADA19986, PhysRevC_70_041604}.
This version was used in our previous work on isoscaling \cite{Fable_PRC_106_024605} and isospin transport \cite{Fable_PRC_107_014604}.
Collisions were followed up to $t_{lim}=300$ fm/c, for a total of approximately $100000$ primary events per system.
Clusters were defined at $t_{lim}$ according to a proximity criterion of the Gaussian centroids in spatial coordinates, with a two-nucleon pair separated by less than 5 fm contributing to the formation of a cluster.
In this version the mean-field description is based on the momentum-dependent Gogny force, consistent with the incompressibility modulus of symmetric matter $K_{sat}=228$ MeV and $\rho_0=0.16$ fm$^{-3}$.
Based on such effective interaction, two parametrizations of the density dependence of the nuclear symmetry energy were exploited, namely the Gogny (soft dependence) and Gogny-AS (stiff dependence) forces \cite{PhysRevC_21_1568}.

Concerning \textsc{amd-cc}, we employed the same simulated data as those published in \cite{CAMAIANI_PRC_102_044607}. 
Collisions were followed up to $t_{lim}=500$ fm/c, for a total of about $40000$ primary events per system.
In this version, the mean-field description is based on the momentum-dependent SLy4 effective Skyrme interaction \cite{CHABANAT1997710}, with $K_{sat}=230$ MeV and $\rho_0=0.16$ fm$^{-3}$.
Two parametrizations of the symmetry energy were tested: a soft symmetry energy dependence with $E_{sym}=32$ MeV and $L=46$ MeV, as zero and first order terms, respectively, and a stiff one with $L=108$ MeV.
Finally, a reduction factor of $\gamma = 0.85$ was employed for the in-medium correction of nucleon-nucleon cross section, following the prescription proposed in \cite{PhysRevC_90_064602}.

For both models, the input impact parameter followed a triangular distribution from $0$ to a value $b_{max}$, that is slightly above the geometrical grazing impact parameter $b_{gr}$ (about $9.7$ fm for $^{40}$Ca$+^{40}$Ca and $10.4$ fm for $^{48}$Ca$+^{48}$Ca).
The statistical decay code \textsc{gemini++} was employed \cite{PhysRevC_82_014610} to deexcite the hot nuclei produced at $t_{lim}$ (afterburner step), using the default parameters of the code, for a total of 50 and 100 secondary events per primary event for \textsc{amd-nc} and \textsc{amd-cc} respectively.
The number of decays per primary event was chosen to limit their effect on the uncertainties associated with the imbalance ratio.

We note an important discrepancy between the two versions of \textsc{amd}, due to the difference in stopping time $t_{lim}$. 
Indeed, in our former work with \textsc{amd-nc} we employed a standard $t_{lim}$ value widely used in the literature with this version of the code \cite{PhysRevC_70_041604}.
In comparison, in the case of \textsc{amd-cc} the authors of ref. \cite{CAMAIANI_PhysRevC_103_014605} ensured that the primary fragments have reached thermodynamical equilibrium (while Coulomb repulsion is negligible) at $t_{lim}$, for the systems under study \cite{PhysRevC_99_064616}.
Given the previous comment, some caveats are to be considered when comparing the primary events of the two model versions.

\subsection{Filter and event selection}\label{subsec_filter}

For both aforementioned versions of the \textsc{amd} model, simulated events were filtered with the same software replica of the experimental setup, to allow comparison between the predictions and the experimental data.

Event detection was simulated within the KaliVeda framework \cite{KaliVeda}.
Concerning VAMOS, the experimental polar and azimuthal angular distribution in the laboratory frame were used to apply cuts on the simulated events, while the energy and identification thresholds of the Si and CsI detectors were considered.
Concerning INDRA, KaliVeda allows a complete description of the detector configuration for the experiment, including geometrical coverage, dead zones, detector resolutions and identification thresholds.
It must be noted that the majority of the simulated events are discarded due to the spectrometer angular acceptance and trigger condition, for a total of about $\approx 75\%$ of the whole statistics.
In this paper we conduct detailed comparisons of the remaining $25\%$ of simulated events to INDRA-VAMOS experimental data.

In addition to the numerical filtering, the same offline selections as in the experiment, described in Sec. \ref{subsec_setup}, were applied to the filtered events.
In summary, once the preliminary selections are applied, the filtered simulated events represent a subset of about of $22\%$ and $18\%$ of the total statistics for the \textsc{amd-cc} and \textsc{amd-nc} models, respectively. 

\subsection{General features}\label{subsec_global results}

In this section we present a comparison of several relevant global observables between the two versions of the \textsc{amd} model, to better apprehend the results on isospin diffusion presented in the next section. 
For the sake of clarity and because we focus on the general features, only the results obtained from a soft symmetry energy are presented at this point of the analysis.  

\subsubsection{Topology of the reaction products}\label{subsubsec_topology}

\begin{figure*}[ht]
\centering
\includegraphics[scale=0.88]{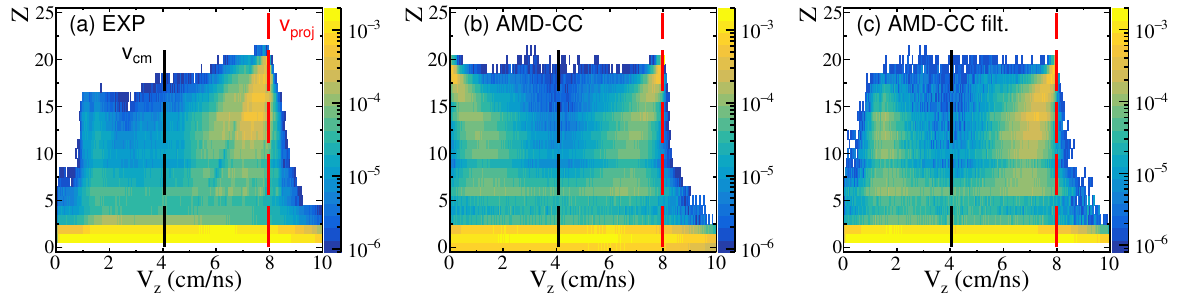}
\caption{Atomic number $Z$ as a function of the parallel velocity $V_{z}$ (in the laboratory frame) for the $^{48}$Ca$+^{48}$Ca system.
Experimental data (a).
Secondary fragments from \textsc{amd-cc} (after \textsc{gemini++} deexcitation) without (b) or with (c) the experimental filter.
The black and red dashed vertical lines indicate the reaction c.m. and the projectile velocities, respectively.
Plots are normalized to the respective total number of events for a better comparison.}
\label{fig_ZVz_exp_mod}
\end{figure*}

Figure \ref{fig_ZVz_exp_mod} depicts the correlations between the charge and the parallel velocity (laboratory frame) for all particles identified in charge in the $^{48}$Ca$+^{48}$Ca reactions, for the experiment (a) and the \textsc{amd-cc} model without (b) and with (c) filter and event selection criteria.
Similarly to our previous work with the \textsc{amd-nc} version \cite{Fable_PRC_106_024605}, the unfiltered model exhibits two main components on both sides of the c.m. velocity (black dashed line), with a third region of LCP and IMF spreading over the whole velocity domain.
By comparing the unfiltered and filtered simulations, we mainly observe the effect of VAMOS angular acceptance, which drastically reduces the measured yields while the right-most component becomes  concentrated in a region of charge and velocity close to the projectile one.
Such topology is representative of dissipative binary collisions, where the fragments detected in VAMOS are mostly the products of the quasiprojectile decay (namely the projectile-like fragment, PLF) resulting from peripheral to semiperipheral collisions, while the products of the quasitarget (target-like fragment, TLF) decay are occasionally identified in INDRA at backward angles (left-most component).
Finally, by comparing the experimental data and the filtered model, a satisfactory agreement is observed. 
It must be noted that similar topologies are observed independently of the version of the \textsc{amd} model and for all the systems, with a similar effect of the numerical filter and data selection \cite{Fable_PRC_106_024605}.

\begin{figure}[h]
\includegraphics[scale=0.4]{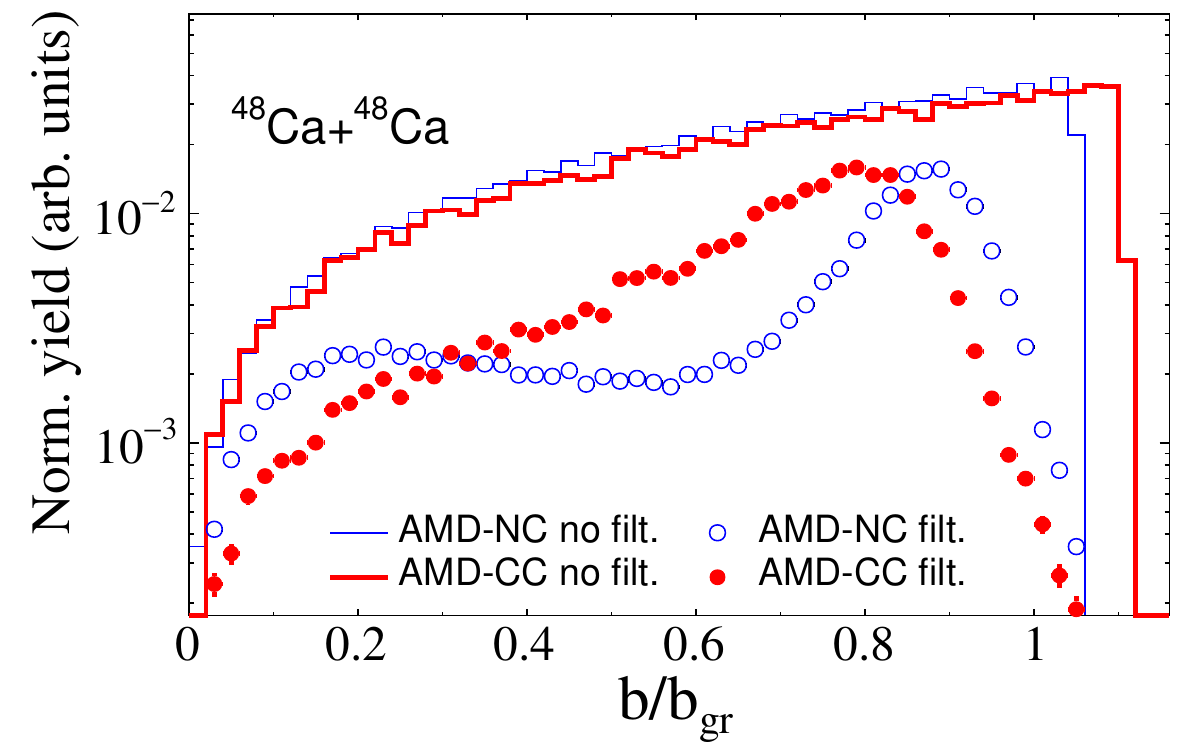}
\caption{Reduced impact parameter distributions from filtered \textsc{amd-nc} (open circles) and \textsc{amd-cc} (full circles) models with data selection.
Solid lines correspond to the input triangular distributions.
The distributions are normalized to the respective total number of simulated events.}
\label{fig_b}
\end{figure}

We present in Fig. \ref{fig_b} the effect of the filter and event selection criteria on the reduced impact parameter distributions for the $^{48}$Ca$+^{48}$Ca system.
The input triangular distributions, represented as solid lines, present a small difference between the models as \textsc{amd-cc} was run up to slightly larger $b/b_{max}$ values than \textsc{amd-nc}.
Interestingly, we observe a strong difference between the two versions of the code once the filter and data selection are applied, \textsc{amd-nc} (open circles) leading to the detection of more peripheral events compared to \textsc{amd-cc} (full circles), even if the latter was run up to higher impact parameter values. 
It must be noted that the exact same results as the ones shown in Fig.\ref{fig_b} (symbols) are obtained when limiting the range of impact parameter with \textsc{amd-cc} to the same range as in \textsc{amd-nc}.
Such remarkable difference is not trivial as it comes from the combined effect of the model dynamics, the detectors acceptance and the applied offline selections. 
In a sense, this plot highlights the inherent difficulty of comparing models with experimental data as the subset of simulated data can vary from one model to another, before and even more after the numerical filtering and data selections are applied.
Indeed, it is for example expected that the \textsc{amd-nc} version leads, on average, to more LCP emissions compared to \textsc{amd-cc}, therefore to a different pattern of multiplicities and kinetic energy spectra.
In fact, a more detailed study of the primary fragments from \textsc{amd-nc} (not presented here) leads to the conclusion that this version of the code tends to overproduce inelastic collisions in the $b/b_{max}=0.8-1$ region as compared to \textsc{amd-cc}, while these fragments have still enough excitation energy to emit LCP in the afterburner phase.
%

\subsubsection{Study of the QP residue}\label{subsubsec_plf}

We present in this section a comparison of the global observables related to the fragment detected in VAMOS, expected to be the QP remnant (PLF).
According to the simulations, more than $95\%$ of the nuclei detected in VAMOS (after data selection criteria) correspond to the remnant of the QP, the latter defined as the biggest primary fragment forward emitted in the c.m. reference frame.

\begin{figure*}[t]
\includegraphics[scale=0.85]{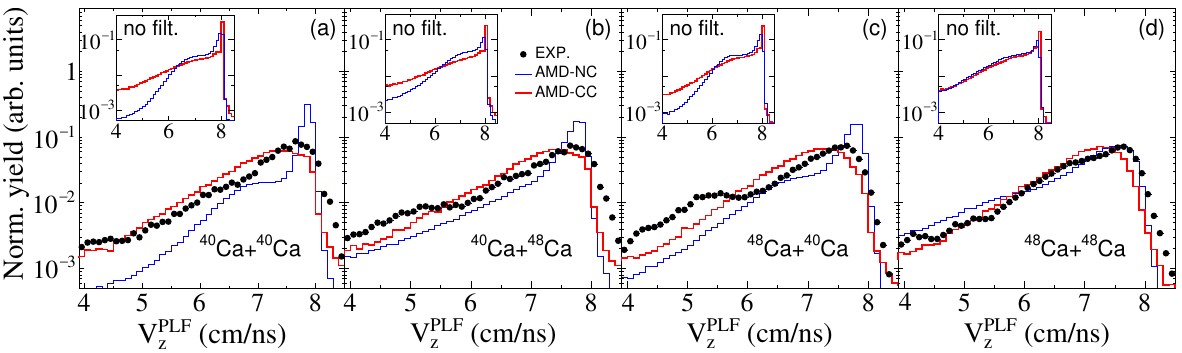}
\includegraphics[scale=0.85]{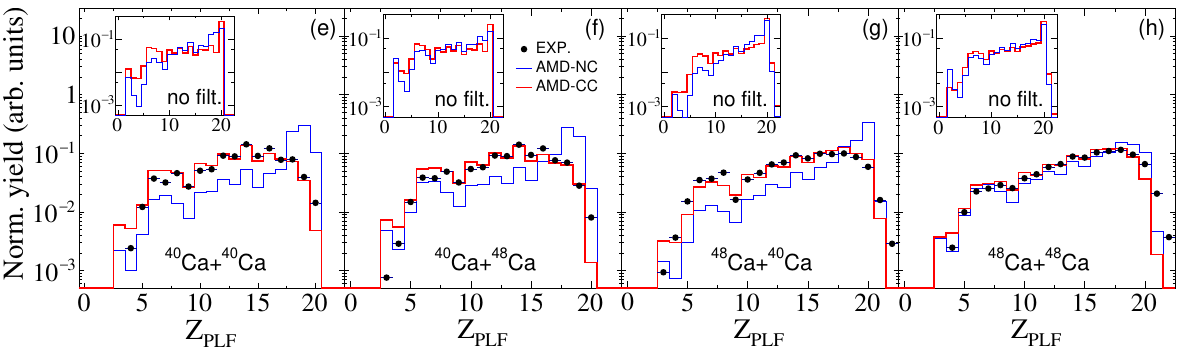}
\caption{Parallel velocity (a)-(d) and charge (e)-(h) distributions of the PLF from the experiment (full circles) and filtered \textsc{amd-nc} and \textsc{amd-cc} models (blue and solid lines, respectively) . 
The inner plots represent the PLF from the models when the numerical filter is not applied.
The distributions are normalized to their integral.}
\label{fig_Vzplf}
\end{figure*} 

The parallel velocity (laboratory frame) and charge distributions of the filtered PLF are given in Figs. \ref{fig_Vzplf}(a)-\ref{fig_Vzplf}(d) and \ref{fig_Vzplf}(e)-\ref{fig_Vzplf}(h) respectively, for the four reactions under study. 
The main plots present the distributions obtained from the experiment (full circles) and filtered models (continuous lines) when the selection criteria are applied.
Concerning the velocity distributions (top panels), we observe decreasing values of the yields with decreasing velocity of the PLF, reflecting the impact parameter distribution.
As expected from Fig. \ref{fig_b}, the differences between the two versions of \textsc{amd} are more pronounced for the most peripheral collisions ($b/b_{gr} \simeq 0.8$, $V_{z} \simeq 8$ cm/ns) as \textsc{amd-nc} tends to overproduce inelastic-like events compared to \textsc{amd-cc}.
It is interesting to note that the opposite effect is observed for the unfiltered models represented in the inner plots, showing that the experimental selection criteria (and indirectly the consideration of cluster formation) may have a strong impact on the filtered distributions.
An overall better agreement with experimental data is obtained with \textsc{amd-cc}, while it tends to produce slighlty slower QP compared to the experiment and \textsc{amd-nc}.
Similar results were obtained from various systems simulated with \textsc{amd-cc} followed by \textsc{gemini++} as afterburner \cite{PhysRevC_103_014603, CAMAIANI_PhysRevC_103_014605, Ciampi_PRC_106_024603}, and it was concluded that the model is more dissipative than the experimental case, although the effect of the filter, event selection (event multiplicity) and clustering process were not detailed. 
Concerning the atomic number distributions (bottom panels), similar conclusions can be drawn: \textsc{amd-cc} reproduces remarkably well the charge distributions for all the systems, even for the neutron-rich $^{48}$Ca$+^{48}$Ca, while \textsc{amd-nc} predicts more heavy products than experimentally observed.
The same conclusion can also be drawn from the mass distributions of the PLF (not shown here).
Finally, it is interesting to note that the observed experimental odd-even staggering \cite{PhysRevC_88_064607}, stronger for neutron-deficient systems, is partially reproduced by both filtered simulations, as a consequence of the sequential decays (\textsc{gemini++}).

\begin{figure}[ht]
\includegraphics[scale=0.42]{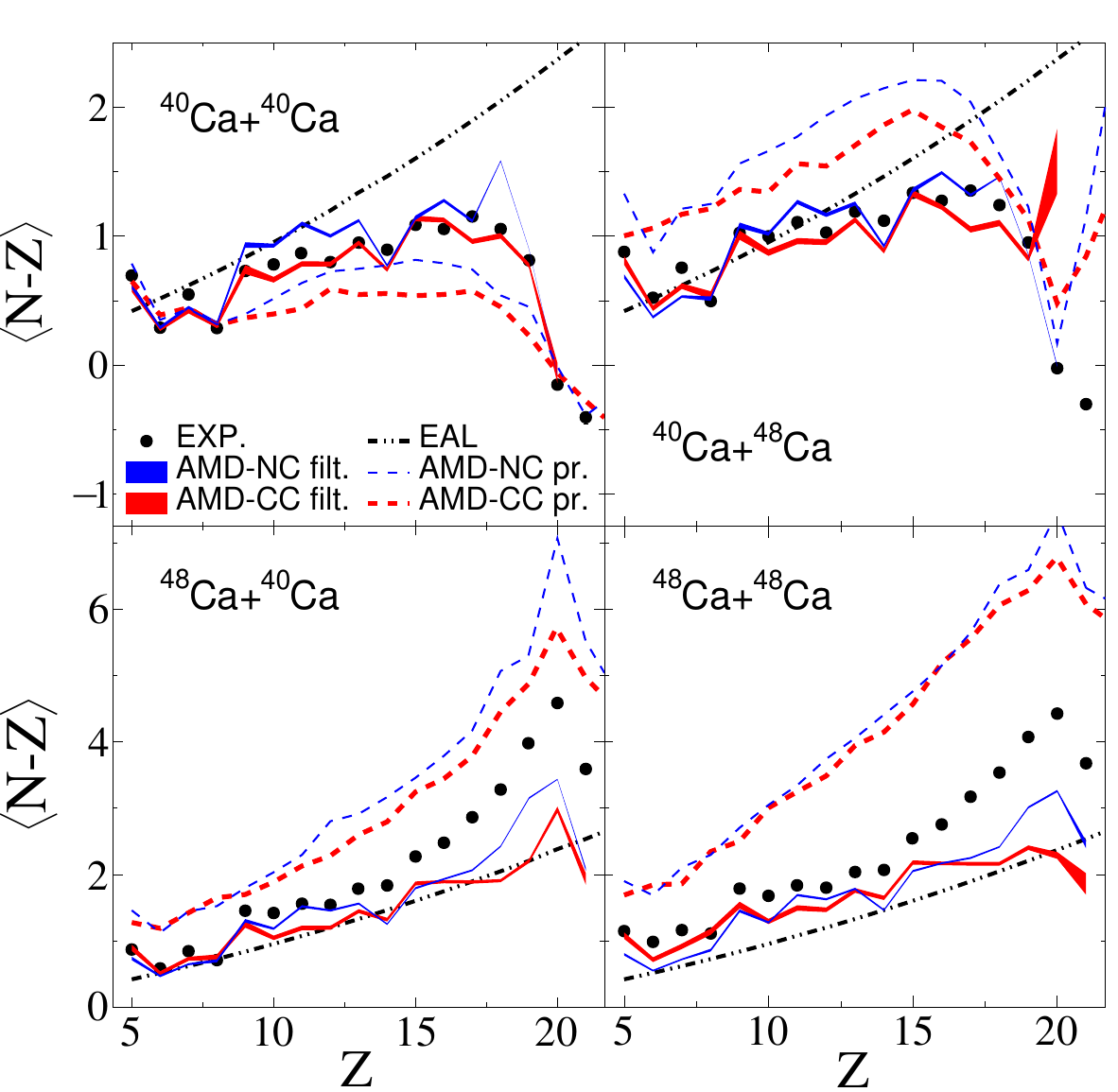}
\caption{Average neutron excess of the PLF for the experimental data (full circles) and filtered \textsc{amd-nc} and \textsc{amd-cc} (blue and red continuous curves, respectively). 
The dashed-dotted line is the EAL from \cite{PhysRevC_58_1073}.
Results from the QP predicted by the model (before \textsc{gemini++}) are plotted in thick red and thin blue dashed lines, respectively.}
\label{fig_NmZplf}
\end{figure}

We present in Fig. \ref{fig_NmZplf} the evolution of the average neutron excess of the PLF detected in VAMOS as a function of its atomic number, for all the reactions.
We first observe that, similarly to the experiment (full cirlces), both models (full lines) present an evolution according to the neutron content of the projectile, progressively driven by the evaporative attractor line (EAL, dashed-dotted line extracted from \cite{PhysRevC_58_1073}) with decreasing charge \cite{Fable_PRC_106_024605}.
Concerning the $^{40}$Ca projectile reactions, we notice a satisfying agreement for both models with the experimental data, except for the most peripheral collisions ($Z=20$) for the neutron-rich target with \textsc{amd-cc}.
Concerning the $^{48}$Ca projectile reactions, we observe that both models tend to underestimate the neutron richness of the PLF. 
This could be ascribed to a too high excitation energy of the QP in both \textsc{amd} versions, leading to the emission of too many neutrons in the afterburner step.
Since model predictions remain too close to the EAL compared to the data, their mutual difference increases with increasing charge. 
We also note more discrepancies between the models in the $Z_{PLF}=15-20$ peripheral region, \textsc{amd-nc} predicting values closer to the experiment compared to \textsc{amd-cc}.
Finally, the neutron excess of the associated QP, as predicted by the model (before \textsc{gemini++}), is also presented in thin and thick dashed lines for \textsc{amd-nc} and \textsc{amd-cc} respectively. 
We remark that the \textsc{amd-nc} model exhibits systematically more neutron-rich QPs than \textsc{amd-cc}.
The neutron excess is partially preserved for the largest PLF ($Z_{PLF}>15$).

\subsubsection{LCP and IMF multiplicities}\label{subsubsec_mult}

In addition to the PLF properties, we present in this section the multiplicity of the nuclei measured in coincidence in INDRA.
We aim at studying the evolution of cluster emission with the dynamical properties of the reactions, using the parallel velocity $V_{z}^{PLF}$ of the fragment detected in VAMOS as a sorting parameter.
Indeed, such a selection allows one to remove the trivial bias induced by the excess of peripheral collisions in \textsc{amd-nc}, as seen in Fig. \ref{fig_Vzplf} in the $V_{z}^{PLF} \simeq 7.5-8$ cm/ns region. 
For this reason, we focus our interpretations on the semiperipheral region ($V_{z}^{PLF} \lesssim 7.25$ cm/ns).

\begin{figure*}[h]
\includegraphics[scale=0.85]{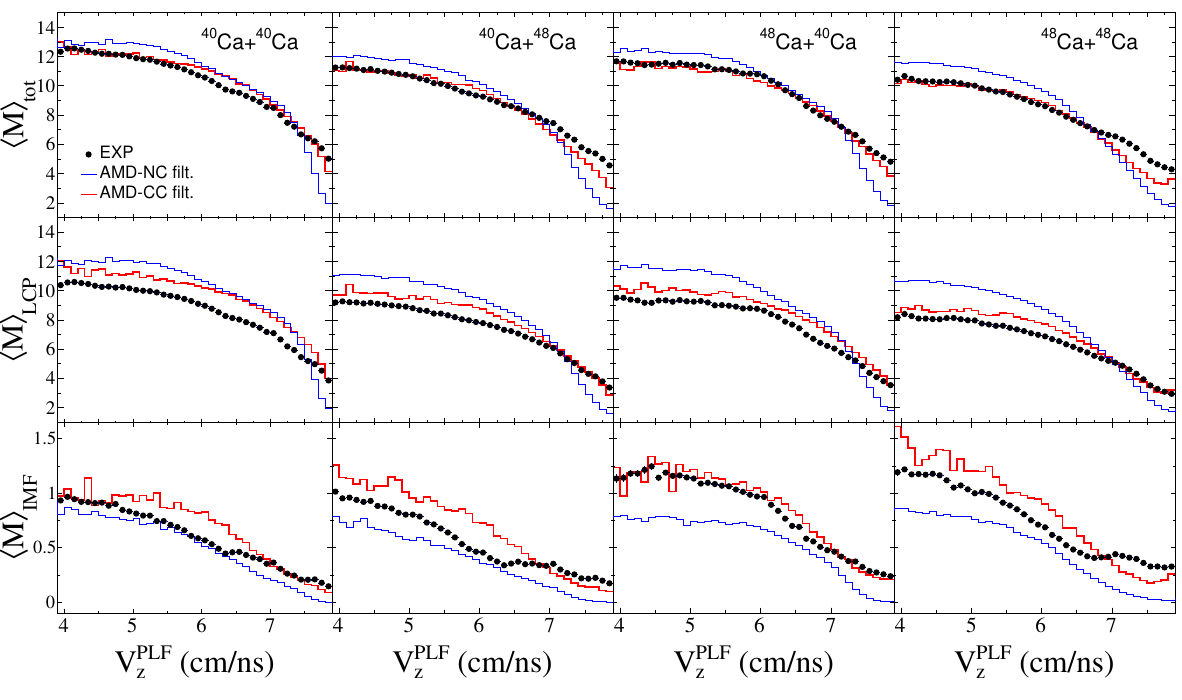}
\caption{Average total (a)-(d), LCP (e)-(h), and IMF (i)-(l) multiplicities measured with INDRA for the experiment (full circles) and the filterd \textsc{amd-nc} (thin blue lines) and \textsc{amd-cc} (thick red lines) simulations followed by \textsc{gemini++}.}
\label{fig_mult}
\end{figure*}

Figure \ref{fig_mult} shows a comparison of the average total, LCP and IMF multiplicity distributions of the nuclei detected in INDRA as a function of $V_{z}^{PLF}$.
We first observe that all multiplicities increase with decreasing $V_{z}^{PLF}$, reflecting an increase of fragment production with increasing dissipation as higher excitation energies are reached on average.
Then, we notice that the total multiplicity is governed by the LCP emissions, which both models tend to overproduce for all $V_{z}^{PLF}$. 	
This overproduction increases with decreasing $V_{z}^{PLF}$, reaching about $\Delta \langle M \rangle_{LCP} \simeq 1$ and $\Delta \langle M \rangle_{LCP} \simeq 2$ for neutron-deficient and neutron-rich projectile reactions, respectively, for \textsc{amd-nc}, while the overproduction remains much smaller for \textsc{amd-cc}. 
A noticeable difference is also observed in the IMF multiplicity distributions that present an underproduction in the case of \textsc{amd-nc} and an overproduction for \textsc{amd-cc}, more evident for the $^{48}$Ca target reactions.

\begin{figure*}[ht]
\includegraphics[scale=0.85]{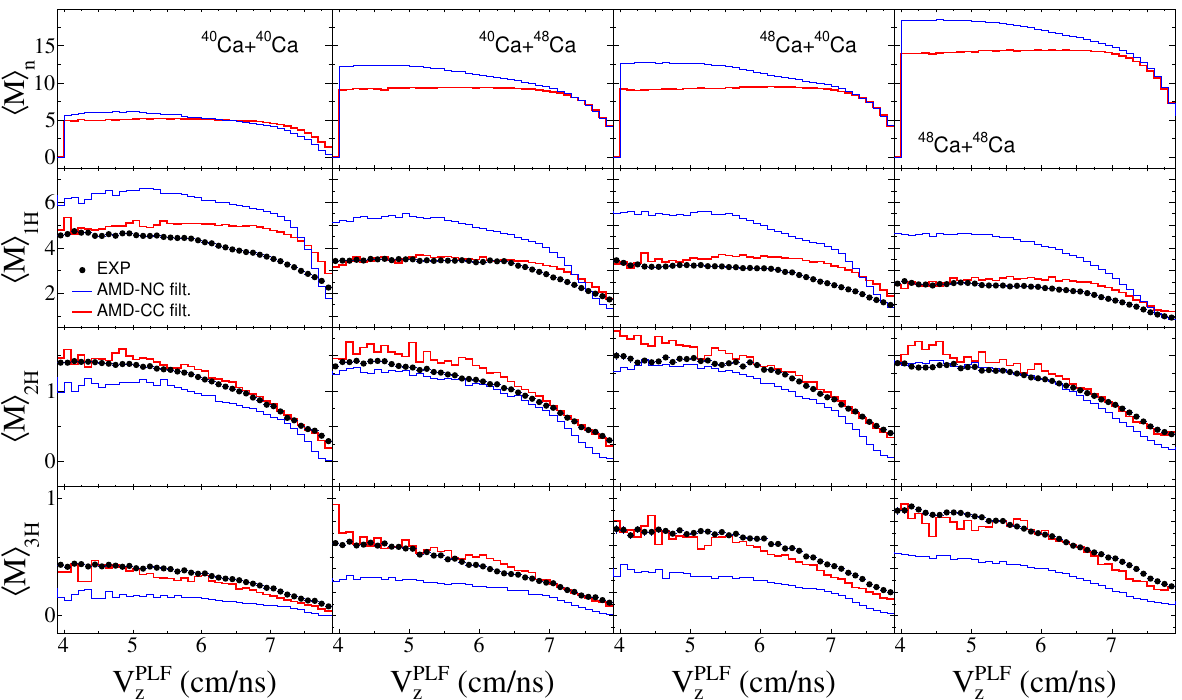}
\caption{Average multiplicities of neutrons and $Z=1$ nuclei (isotopically identified) for the experiment (full circles) and the filterd \textsc{amd-nc} (thin blue lines) and \textsc{amd-cc} (thick red lines) simulations followed by \textsc{gemini++}. 
The neutron distributions are not filtered.}
\label{fig_Z1_mult}
\end{figure*}

\begin{figure*}[ht]
\includegraphics[scale=0.85]{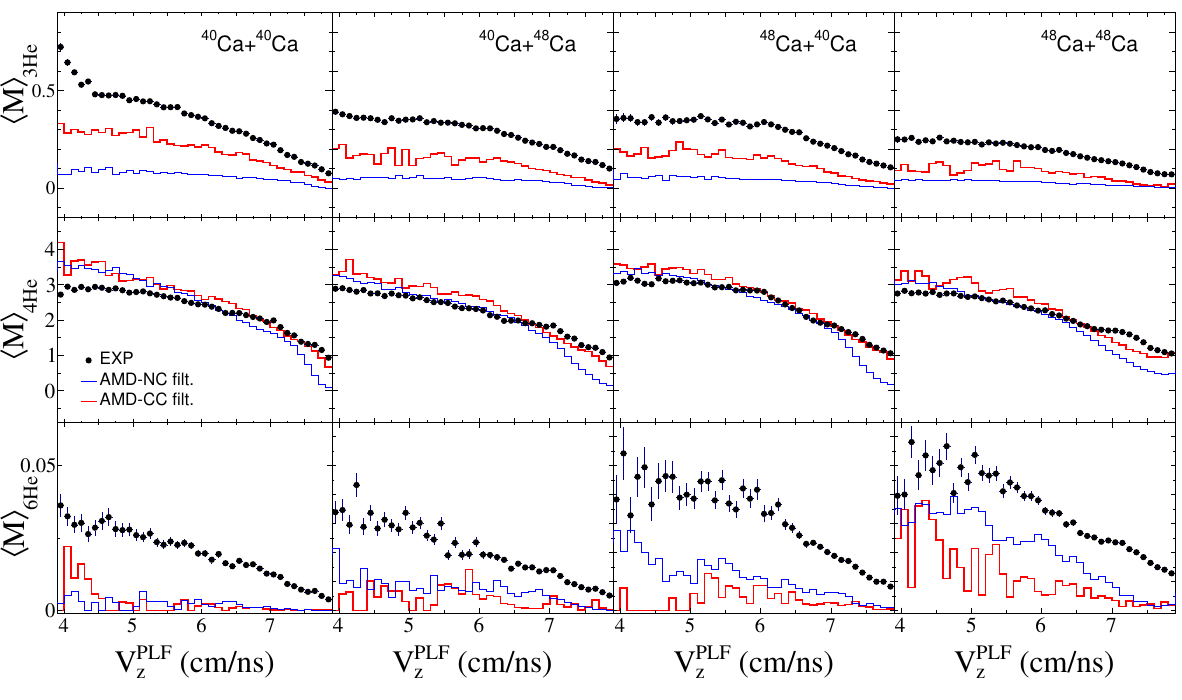}
\caption{Same as Fig. \ref{fig_Z1_mult} for $Z=2$ nuclei.}
\label{fig_Z2_mult}
\end{figure*}

We present in Figs. \ref{fig_Z1_mult} and \ref{fig_Z2_mult} the average multiplicities of $Z=1$ and $Z=2$ nuclei isotopically identified in INDRA as a function of $V_{z}^{PLF}$. 
The unfiltered neutron distributions are also shown in Fig. \ref{fig_Z1_mult}.
As a first general comment, we observe that the models exhibit a dominant emission of proton and $\alpha$ particles, similarly to the data.
Concerning the $Z=1$ isotopes, we observe an overproduction of protons by \textsc{amd-nc}, reaching about $\approx 40\%$ for the most dissipative reactions, while \textsc{amd-cc} presents values close to the experimental one at low velocity.
Interestingly, an overproduction of neutrons by \textsc{amd-nc} compared to \textsc{amd-cc} is also observed for all the systems except the neutron-deficient $^{40}$Ca$+^{40}$Ca.
This indicates that in the case of \textsc{amd-nc} more free protons (and neutrons) are produced to the detriment of clusters, while the difference in filtered total multiplicity remains reasonable.
This seems to be confirmed by the $^{2,3}$H isotopes multiplicities, which are systematically underproduced by \textsc{amd-nc}, while the inclusion of cluster formation in \textsc{amd-cc} leads to a better agreement with the data. 
Nevertheless, one should exercise caution regarding the $^{2}$H isotopes because the SLy4 effective force employed in \textsc{amd-cc} leads to an overestimation of their binding energy for the soft parametrization, leading to an anomalous increase of their multiplicity \cite{PhysRevC_99_064616}.

Concerning now the $Z=2$ isotopes, we first notice an underproduction of $^{3}$He isotopes from the models, reaching a reduction of about $75\%$ and $50\%$ for \textsc{amd-nc} and \textsc{amd-cc}, respectively.
The situation is even worse for $^{6}$He isotope multiplicities, even if the average multiplicities are very small.   
Finally, a similar behavior is observed in the case of $\alpha$ particles for both models, without clear evidence of what model performs the best.   

\subsubsection{Transverse kinetic energy of LCP}\label{subsubsec_Et12}

\begin{figure*}[ht]
\includegraphics[scale=0.85]{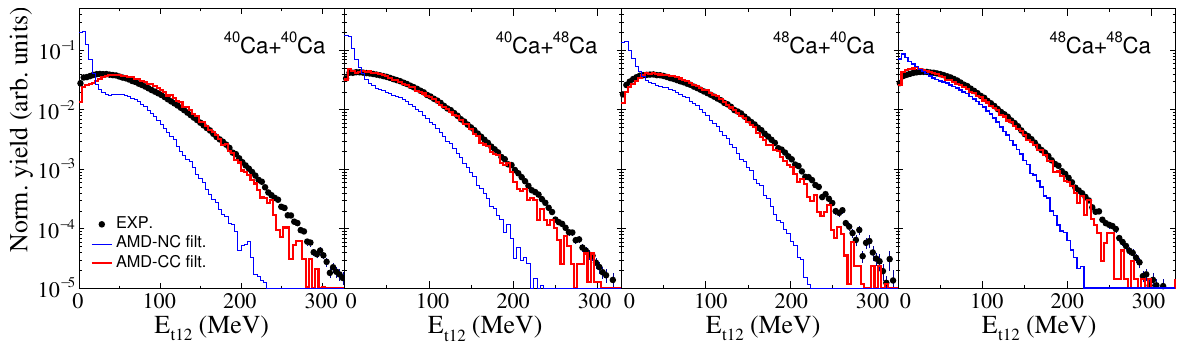}
\caption{Total transverse kinetic energy distributions of light charged particles (normalized to their integral) for the experiment (full circles) and the filtered \textsc{amd-nc} (thin blue lines) and \textsc{amd-cc} (thick red lines) simulations followed by \textsc{gemini++}.}
\label{fig_Et12}
\end{figure*}

We present in Fig. \ref{fig_Et12} the distributions of the total transverse energy of the LCP identified in charge, namely $E_{t12}$, extracted from the filtered models and the experiment.

This global variable is of particular interest for centrality estimation in the analysis of isospin transport (see Sec. \ref{sec_isodiff}) as it is well suited to the performance of the INDRA array for which LCP are detected with a $90\%$ efficiency.
It is defined as 
\begin{equation}
E_{t12}=\sum_{i:Z_{i}\leq2}E_{i}\sin^{2}\theta_{i}
\label{eq_et12} 
\end{equation}
where in the sum $i$ runs over the detected (filtered) products of each event with $Z_{i}\leq2$, laboratory kinetic energy $E_{i}$ and laboratory polar angle $\theta_{i}$.

We first notice in Fig. \ref{fig_Et12} a good agreement between the experimental distributions (full circles) and the filtered \textsc{amd-cc} simulations (thick red lines) that reproduce both the experimental trends and $E_{t12}$ values, including the tail of the distributions.
In the case of \textsc{amd-nc}, we observe more narrowed-down distributions with higher statistics for the less dissipative collisions ($E_{t12} \lesssim 20 $ MeV), as already anticipated from Fig. \ref{fig_Vzplf}, while the tails of the experimental data are not reproduced.
We clearly observe in this figure the relevance of considering clustering to reproduce the experimental kinetic properties of the LCP, as recently evidenced in \cite{PhysRevC_107_044614}. 

\subsection{Discussion}\label{subsec_model_discussion}

As a general comment, we can conclude that both versions of \textsc{amd} reproduce reasonably well the experimental topology (charge and parallel velocity correlations), highlighting the effect of VAMOS acceptance (and trigger) to favor the measurement of the QP remnant in binary dissipative collisions.
A comparison of the filtered simulation impact parameter distributions shows that the data selection criteria have a strong impact on the subset of simulated data from one model to another.    

We have shown that the \textsc{amd-cc} version reproduces remarkably well the experimental velocity, charge and mass distributions of the PLF detected in VAMOS, while a sizable disagreement is obtained with \textsc{amd-nc}, mostly for the less dissipative collisions. 
This was anticipated as the \textsc{amd-nc} version leads to an overproduction of inelastic-like events compared to \textsc{amd-cc}.

Furthermore, a focus on the evolution of the neutron richness of the PLF shows that both models manage to reproduce, to some extent, the experimental trends.
This point is particularly relevant for the study of isospin diffusion, as it is expected to be responsible of such an evolution according to the neutron content of the projectile.
A significant discrepancy between the models and the experiment is nonetheless observed for the neutron-rich $^{48}$Ca projectile reactions. In our understanding, this disagreement can mainly be ascribed to a too high excitation energy obtained from both \textsc{amd} models and inputted in the afterburner.
Indeed, by comparing the average excitation energy per nucleon obtained from the reconstructed quasiprojectile for the experiment and the models, we estimate a systematic overestimation of about $\simeq 1$ MeV/nucleon from \textsc{amd-cc} and even more from \textsc{amd-nc}.
Consequently, we expect that too many neutrons are emitted by the neutron-rich QPs in \textsc{amd}+\textsc{gemini++} calculations, compared to what can be expected from the experiment.
Also, as the same version of \textsc{gemini++} was employed and both models present a similar evolution of the QP asymmetry for $Z<20$ (thick red and thin blue dashed lines for the $^{48}$Ca projectile reactions in Fig.\ref{fig_NmZplf}), an inaccuracy in the level density parameter in the afterburner is not to be excluded.
Such discrepancy could arise from the neutron-to-proton asymmetry and the excitation-energy dependence of the level-density parameter encoded in \textsc{gemini++}.
It is anticipated that the former prevails for light nuclei very close to the neutron and proton drip lines while the latter is expected for heavy nuclei ($A \gtrsim 120$) \cite{PhysRevC_71_024310, PhysRevC_82_014610}. However, both scenarios fall outside the scope of the current analysis.

By exploiting the parallel velocity of the PLF as a surrogate for the collision dissipation, we have shown that it is also possible to discuss several aspects of the dynamical emission of clusters.
More specifically, comparisons of the average multiplicities have stressed that the inclusion of cluster correlations helps to better reproduce the experimental isotopic multiplicities. 
A remarkable agreement is obtained for $^{1,2,3}$H isotopes in the case of \textsc{amd-cc}, while \textsc{amd-nc} tends to overproduce free protons to the detriment of clusters.
This is nonetheless less obvious for $^{3,4,6}$He isotopes, as both models tend to underproduce $^{3,6}$He but present a satisfying agreement for $^{4}$He.
It was also shown that \textsc{amd-cc} tends to better reproduce the average IMF multiplicties.

To conclude, similarly to \cite{PhysRevC_96_034622}, this work demonstrates that \textsc{amd-cc} is relevant not only in the study of central collisions, but also for semiperipheral to peripheral collisions, where binary dissipative collisions exhaust a large part of the total reaction cross section.
We also showed that the inclusion of cluster formation is a mandatory step to better reproduce the experimental features, such as the multiplicity and transverse kinetic energy distributions.
This is particularly interesting in order to improve the comparison protocols used in the study of HIC, as these global observables are used for experimental impact parameter sorting. 
  
\section{Impact parameter reconstruction}\label{sec_ip_recon}

In this section, we apply a method for impact parameter distribution estimation \cite{PhysRevC_97_014905, PhysRevC_98_024902}, recently adapted from relativistic HICs to the Fermi energy domain by Frankland \textit{et~al.} \cite{PhysRevC_104_034609}.
The goal of the method is to infer realistic impact parameters from the inclusive distribution of an experimentally measured observable.
Such a method is mandatory for suitable comparisons with dynamical models where, as shown in the previous section, we can expect a strong variation in the inclusive impact parameter distribution.
More details of the implementation of the method for the INDRA-VAMOS data are given in Appendix \ref{app1_ip}.

We would like to highlight that the software necessary to perform the present analysis is currently implemented in the KaliVeda heavy-ion analysis toolkit \cite{KaliVeda}.

\subsection{Validation of the method with \textsc{amd-cc}}\label{subsec_ip_model}

In the following, we present the results of applying the impact parameter reconstruction method presented in Appendix \ref{app1_ip} to the filtered \textsc{amd-cc} simulated data.
Since, as discussed in Sec. \ref{sec_simu}, \textsc{amd-cc} reproduces several experimental findings much better than the \textsc{amd-nc} version, we will now exclusively use and focus on the former. 

\begin{figure}[ht]
\centering
\includegraphics[scale=0.42]{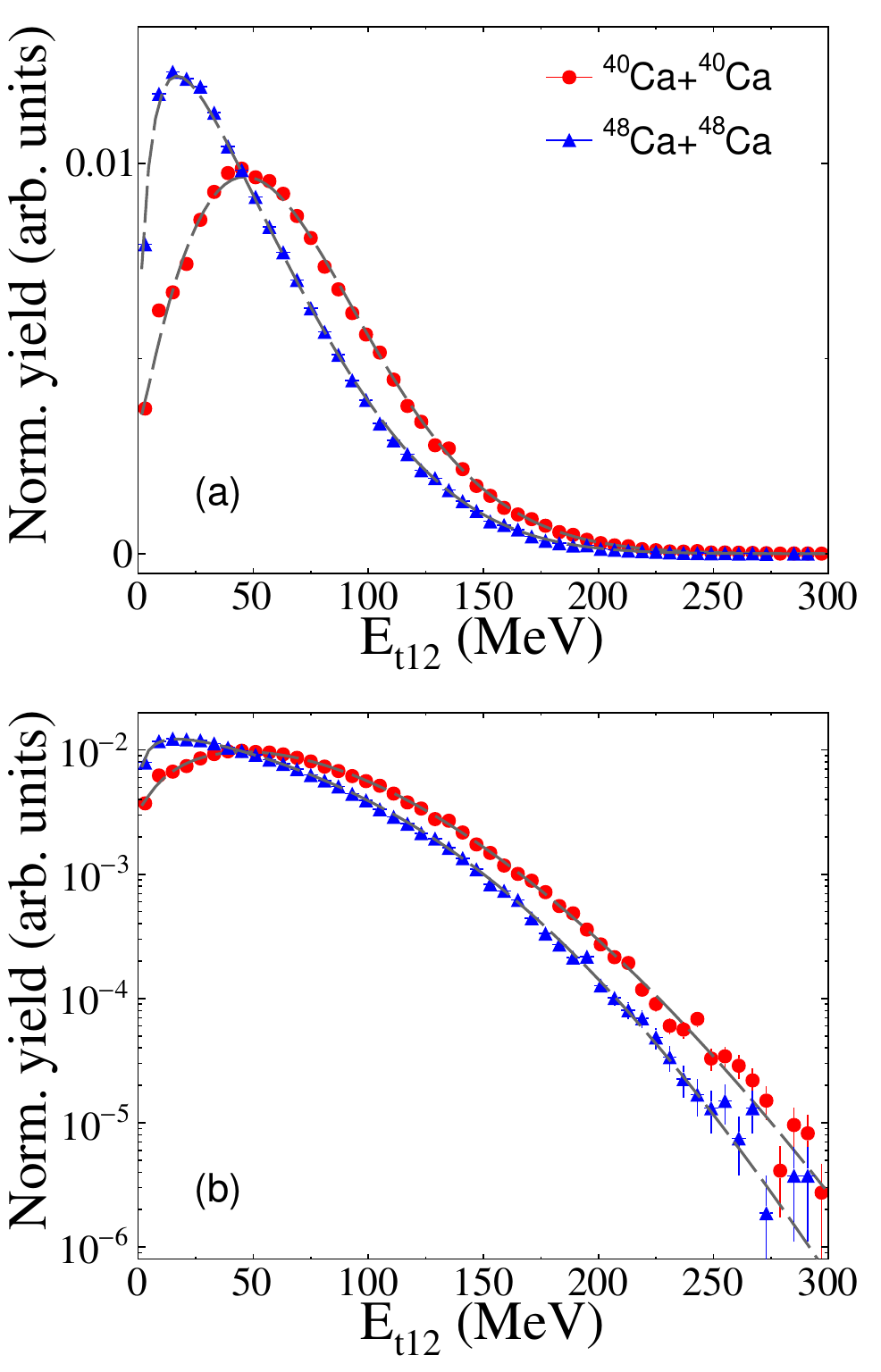}
\caption{Model calculations: fits (dashed curves) to the inclusive distribution of $E_{t12}$ from \textsc{amd-cc} (stiff) filtered model predictions (followed by \textsc{gemini++}), for $^{40}$Ca$+^{40}$Ca and $^{48}$Ca$+^{48}$Ca reactions. 
Each distribution is presented in both (a) linear and (b) logarithmic $y$ axes. 
Statistical uncertainties are not shown when smaller than the symbols.}
\label{fig_et12_mod_fits}
\end{figure}
Figure \ref{fig_et12_mod_fits} shows an example of the quality of the fits to the inclusive $E_{t12}$ data histograms from \textsc{amd-cc} stiff, achieved using Eq. \ref{eq_PX_inclusive} and the gamma-distribution parametrization.
The fit parameters and the reduced $\chi^2$ values are given in Table \ref{tab_et12_mod_fits} for all simulated data and both parametrizations of the symmetry energy.
We observe that the shapes of the filtered distributions are globally well reproduced by the fits, with a satisfactory goodness-of-fit parameter ($\chi^{2} \approx 1$) and similar parameter values for both parametrizations (for a given system).
\begin{table}[ht]
\centering
	\begin{tabular}{c|c|c c c c c c c}
    Model & System & $\alpha$ & $\gamma$ & $\theta$  & $X_{min}$  & $X_{max}$ & $\chi^{2}$ \\
          &        &          &          & [MeV]     & [MeV]      & [MeV]     &       \\       
	\hline
    \multirow{4}{*}{\shortstack{\textsc{amd-cc} \\ stiff}} & $^{40}$Ca$+^{40}$Ca & 0.10 & 0.52 & 8.95  & 1 & 265 & 1.2 \\
                              & $^{40}$Ca$+^{48}$Ca & 0.26 & 0.81 & 9.42  & 4 & 197 & 1.7 \\   
                              & $^{48}$Ca$+^{40}$Ca & 0.12 & 0.59 & 8.68  & 5 & 251 & 1.4 \\
                              & $^{48}$Ca$+^{48}$Ca & 0.33 & 0.93 & 8.48  & 8 & 179 & 1.1 \\
    \hline
    \multirow{4}{*}{\shortstack{\textsc{amd-cc} \\ soft}} & $^{40}$Ca$+^{40}$Ca & 0.15 & 0.58 & 7.12  & 1  & 246 & 1.5 \\
                              & $^{40}$Ca$+^{48}$Ca & 0.31 & 0.84 & 9.65  & 5  & 187 & 1.2 \\   
                              & $^{48}$Ca$+^{40}$Ca & 0.15 & 0.60 & 8.71  & 5  & 241 & 1.3 \\
                              & $^{48}$Ca$+^{48}$Ca & 0.35 & 0.96 & 9.68  & 10 & 175 & 1.1 \\  		
	\end{tabular}
\caption{Model calculations: results of the fits to the total transverse energy. $\chi^{2}$ is the reduced chi-square value of each fit.}
\label{tab_et12_mod_fits}	
\end{table}
\begin{figure*}[ht]
\includegraphics[scale=0.65]{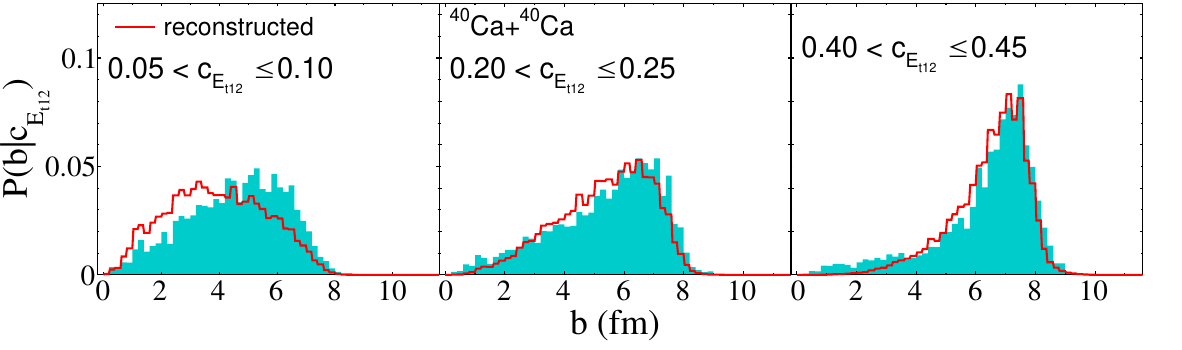}
\includegraphics[scale=0.65]{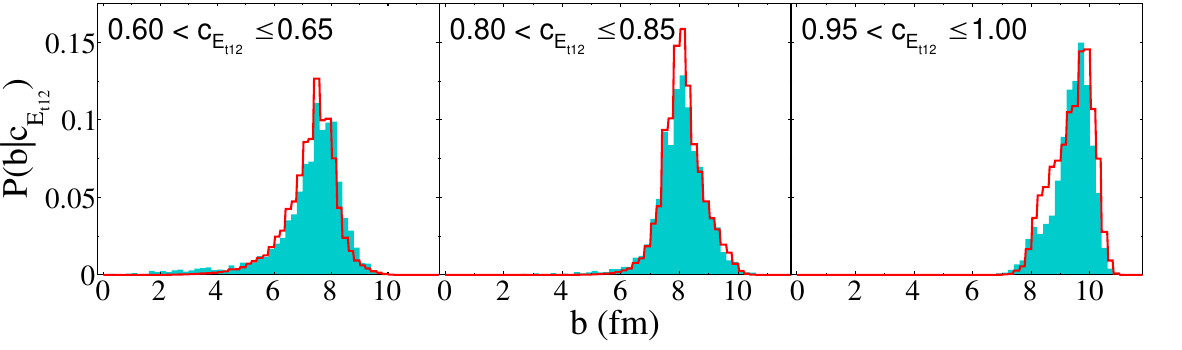}
\caption{Model calculations: impact parameter probability distributions for various samplings of experimental centrality $c_{E_{t12}}$, returned by the fit (solid lines) and calculated directly (shaded area), for the same $^{40}$Ca$+^{40}$Ca \textsc{amd-cc} (stiff) filtered events as in Fig \ref{fig_et12_mod_fits}.}
\label{fig_breal_besti}
\end{figure*}

Fitting the $P(E_{t12})$ distributions allows to determine the conditional probability distribution $P(E_{t12}|c_{b})$, which is then used to extract the centrality and absolute impact parameter distributions for any sample $\mathbb{S}$ (see Eqs. \ref{eq_PCb_for_S} and \ref{eq_Pb_for_S}).
For this analysis, we have adopted a sampling of 20 bins of $5\%$ experimental centrality $c_{E_{t12}}$.

Results for the $^{40}$Ca$+^{40}$Ca collisions are shown in Fig. \ref{fig_breal_besti} for the \textsc{amd-cc} model followed by \textsc{gemini++} as afterburner, for six centrality intervals.
The distributions reconstructed from the deduced form of $P(E_{t12}|c_{b})$ (solid lines) are compared to the actual distributions (filled areas), directly obtained by applying the corresponding cuts in $c_{E_{t12}}$ to the model. 
Also, the average impact parameter expected to be probed in the most central collisions is about $\langle b \rangle \simeq 3 \pm 0.5$ fm, corresponding to semicentral collisions. 
This limit is mostly induced by the VAMOS trigger condition, preventing measurement of the most central collisions.
Similar results are obtained from \textsc{amd-cc} with a soft symmetry energy. 

We observe a reasonable agreement between the distributions for all the samplings, with similar results for all the simulated systems.
Also, similarly to \cite{PhysRevC_104_034609}, the most central sampling ($0.05 < c_{E_{12}} \leq 0.1$) presents a shift that would lead to a slight underestimation of the mean impact parameter.

\begin{figure}[]
\centering
\includegraphics[scale=0.42]{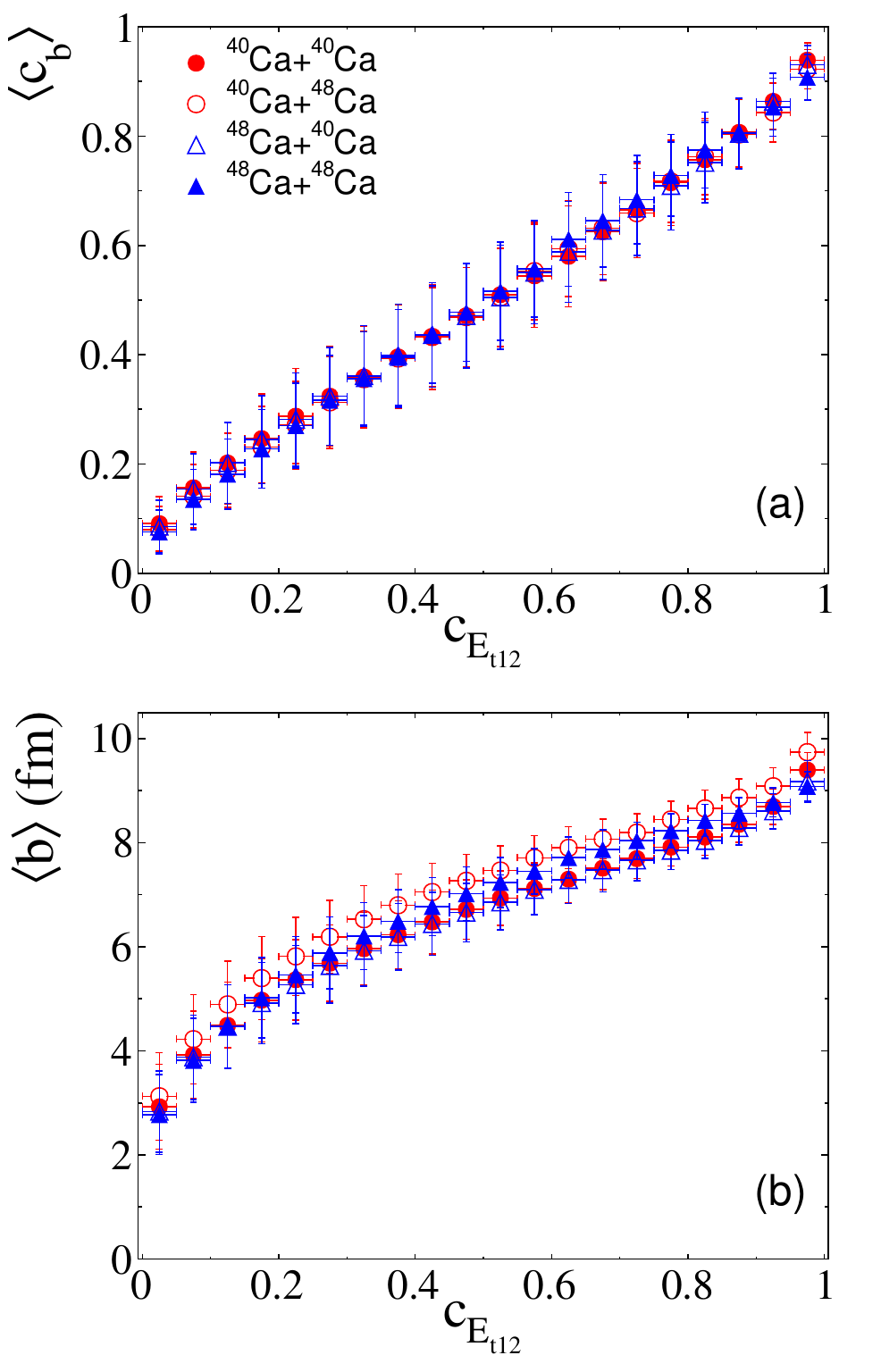}
\caption{Model calculations: evolution of the average $b$ centrality (a) and impact parameter (b) with the applied sampling of $5\%$ in $c_{E_{t12}}$, for the \textsc{amd-cc} (stiff) filtered simulations (followed by \textsc{gemini++}).
The error bars correspond to the standard deviation.}
\label{fig_meancb_meanb_vs_cet12_mod}
\end{figure}

Finally, we present in Fig. \ref{fig_meancb_meanb_vs_cet12_mod} the evolution of the mean centrality $\langle c_{b} \rangle$ and impact parameter $\langle b \rangle$ as a function of the applied sampling in $c_{E_{t12}}$, for all the considered systems. The associated standard deviation are shown as error bars.

This plot illustrates the performances of the method to quantitatively characterize the centrality of selected event samples, in a model-independent way (at least for $c_{b}$).
We observe a linear correlation between the reconstructed $c_{b}$ and $c_{E_{t12}}$ in the $\langle c_{b} \rangle \simeq 0.05-0.95$ range, independently of the system.
This evolution is not observed for the reconstructed $b$, where values present a system dependence and increase nonlinearly from $\langle b \rangle \simeq 3$ fm to $\langle b \rangle \simeq 9 $ fm.

\subsection{Application to the experimental data}\label{subsec_ip_exp} 

Following the same procedure as in the previous section, we can estimate the impact parameter distributions for the experimental data.

\begin{figure}[ht]
\centering
\includegraphics[scale=0.42]{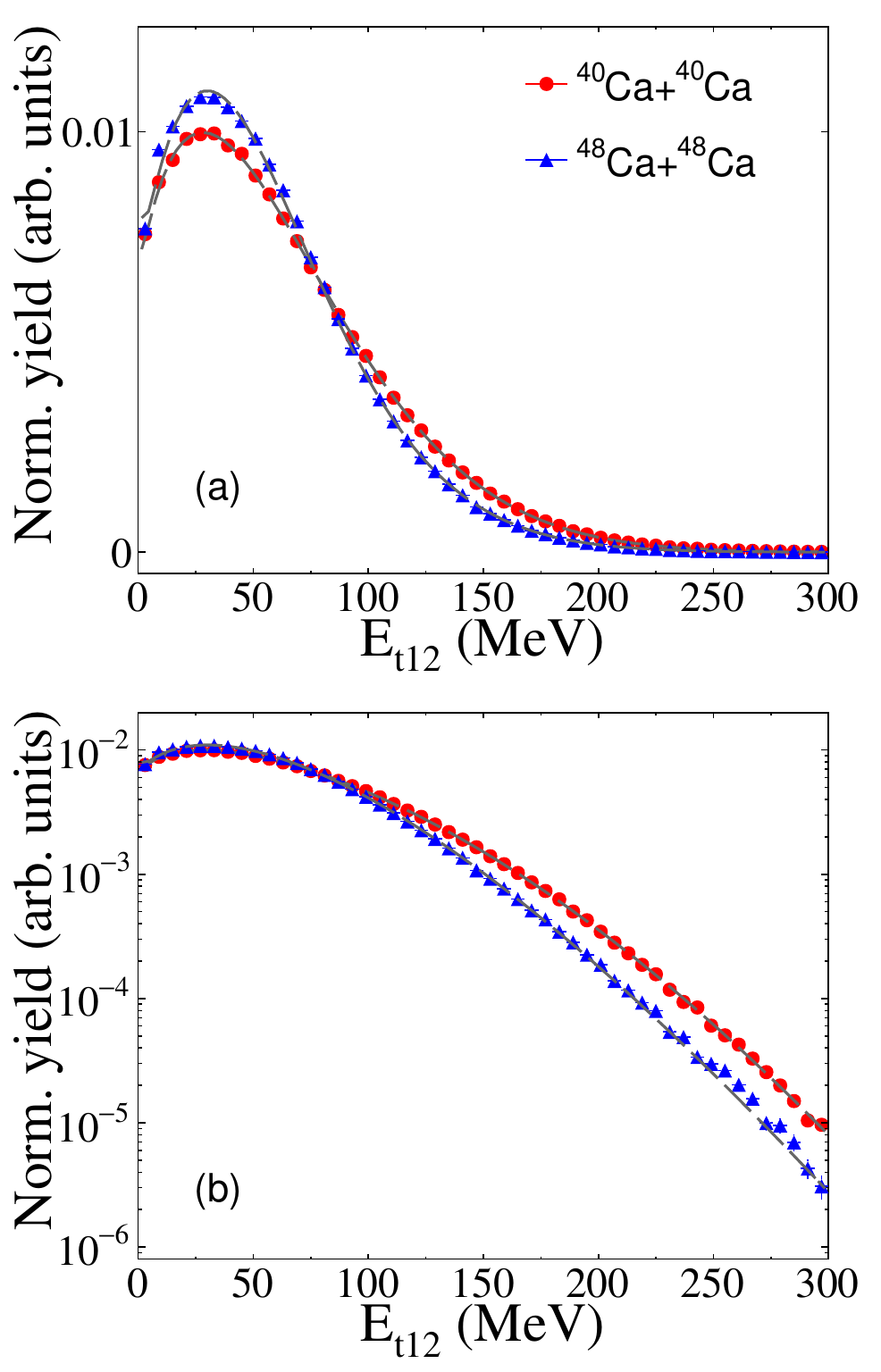}
\caption{Experimental data: fits (dashed curves) to the inclusive distribution of $E_{t12}$ for the $^{40}$Ca$+^{40}$Ca and $^{48}$Ca$+^{48}$Ca reactions. 
Each distribution is presented in both (a) linear and (b) logarithmic $y$ axes. 
Statistical uncertainties are not shown when smaller than the symbols.}
\label{fig_et12_exp_fits}
\end{figure}
\begin{figure}[ht]
\centering
\includegraphics[scale=0.42]{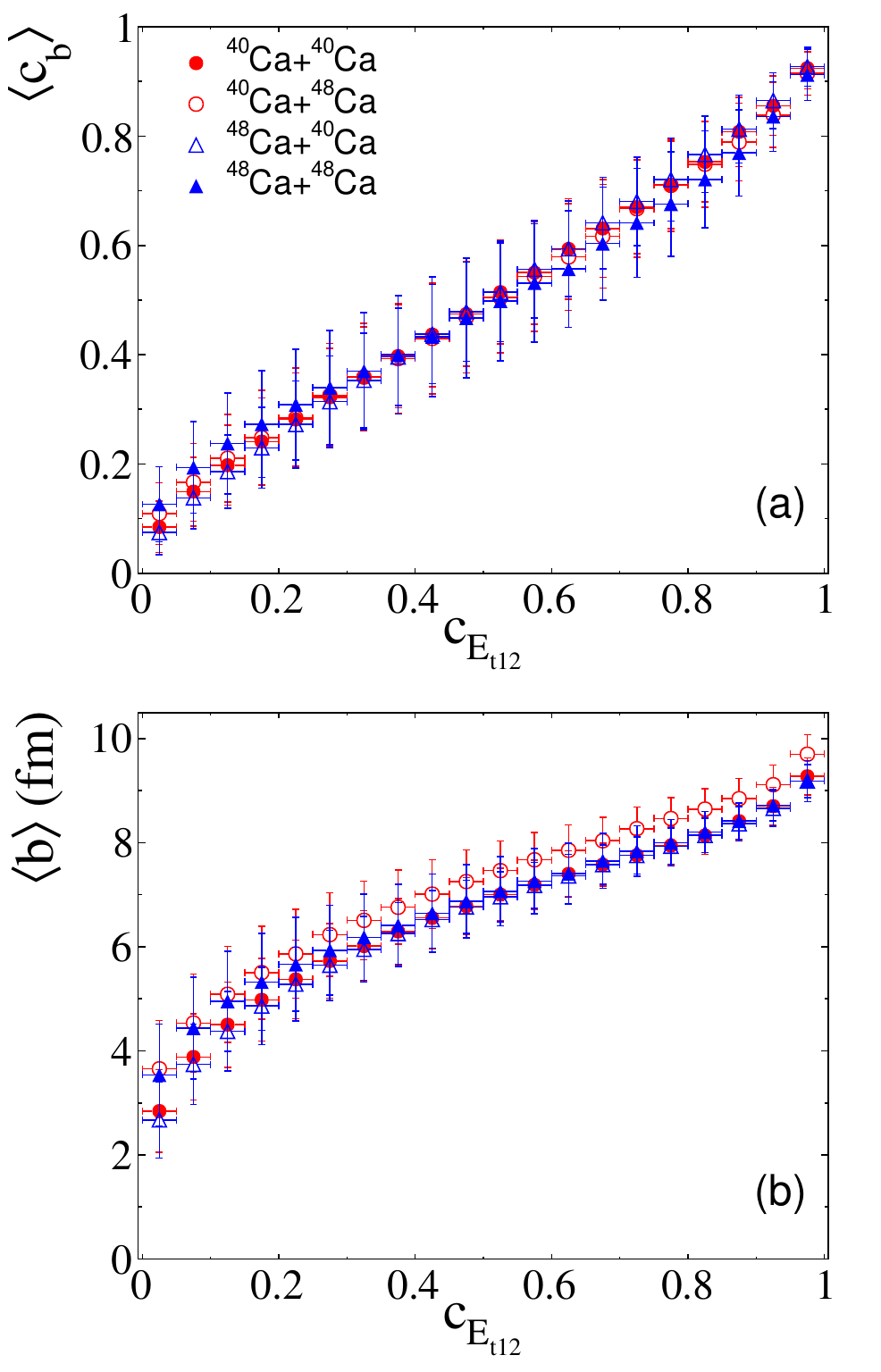}
\caption{Experimental data: evolution of the average $b$ centrality (a) and impact parameter (b) with the applied sampling of $5\%$ in $c_{E_{t12}}$.
The error bars correspond to the standard deviation.}
\label{fig_meancb_meanb_vs_cet12_exp}
\end{figure}

Figure \ref{fig_et12_exp_fits} shows an example of the quality of the fits to the inclusive $E_{t12}$ experimental data, achieved using Eq. \ref{eq_PX_inclusive} and the gamma-distribution parametrization, similarly to Fig. \ref{fig_et12_mod_fits}.
The fit parameters and the reduced $\chi^2$ values are given in Table \ref{tab_et12_exp_fits} for the four systems under study.
The centrality and absolute impact parameter distributions for all the samples in experimental centrality $c_{E_{t12}}$ are then estimated using Eqs. \ref{eq_PCb_for_S} and \ref{eq_Pb_for_S}, respectively.
\begin{table}[H]
\centering
	\begin{tabular}{c|c c c c c c c}
    System & $\alpha$ & $\gamma$ & $\theta$  & $X_{min}$  & $X_{max}$ & $\chi^{2}$ \\
           &          &          & [MeV]     & [MeV]      & [MeV]     &            \\       
	\hline
    $^{40}$Ca$+^{40}$Ca & 0.14 & 0.69 & 10.29 & 4 & 278 & 1.3 \\
    $^{40}$Ca$+^{48}$Ca & 0.37 & 0.82 & 12.90 & 7 & 167 & 1.3 \\   
    $^{48}$Ca$+^{40}$Ca & 0.17 & 0.71 & 9.11  & 6 & 257 & 1.1 \\
    $^{48}$Ca$+^{48}$Ca & 0.10 & 0.55 & 13.01 & 1 & 233 & 1.4 \\ 		
	\end{tabular}
\caption{Experimental data: results of the fits to the total transverse energy. $\chi^{2}$ is the reduced chi-square value of each fit.}
\label{tab_et12_exp_fits}	
\end{table}
The evolution of the average centrality $\langle c_{b} \rangle$ and impact parameter $\langle b \rangle$ as a function of $c_{E_{t12}}$ are presented in Figs. \ref{fig_meancb_meanb_vs_cet12_exp}(a) and \ref{fig_meancb_meanb_vs_cet12_exp}(b), respectively, for the four reactions under study.
For both observables we find, as expected, increasing values with increasing $c_{E_{t12}}$, even if a stronger system-dependence than the model is observed for $\langle b \rangle$. 
Concerning $\langle b \rangle$, we have used the inclusive impact parameter of the filtered model (after data selection criteria) as a surrogate for $P(b)$ in Eq. \ref{eq_Pb_for_S}, similarly to Sec. \ref{subsec_ip_model}.
The plots given in Fig. \ref{fig_meancb_meanb_vs_cet12_exp}(b) are thus obtained with the inclusive impact parameter distributions extracted from the filtered \textsc{amd-cc} with a stiff symmetry energy.
It was verified that the same quantitative results are obtained with the soft symmetry energy (within the error bars), as the $P(b)$ distributions are almost not sensitive to the employed parametrization.  

\section{Isospin diffusion}\label{sec_isodiff}

This section is dedicated to the study of isospin diffusion from the quasiprojectile (QP) and its remnant (PLF), defined as the fragment measured in VAMOS.
The isospin transport ratio, described in the following, is used to quantify the degree of isospin equilibration, while its evolution is followed as a function of the centrality or impact parameter, estimated using the method described in Sec. \ref{sec_ip_recon}.
In the first part, we focus on the \textsc{amd-cc} model, in order to characterize the effect of the impact parameter reconstruction on the isospin equilibration measured from the QP along with the effect of sequential decays.
In continuity with our previous works \cite{Fable_PRC_106_024605, Fable_PRC_107_014604}, the experimental method applied to reconstruct the primary QP fragments, including the estimation of evaporated neutrons, is also tested and compared to the result obtained from the actual QP predicted by the model. 
Details about the method are given in Appendix \ref{app2_Mn}.
In the second part, the same protocol will be applied to the experimental data, allowing direct comparisons with the model.
  
\subsection{Isospin transport ratio}\label{subsec_ratio}	

The isospin transport ratio (also called imbalance ratio) was introduced by Rami \textit{et~al.} to deduce quantitative signals of isospin diffusion in experimental data \cite{Rami_PhysRevLett_84_1120}.
It consists of combining an isospin-sensitive observable, measured under the same experimental conditions, with four systems differing in their initial neutron-to-proton ratios.
It is defined as
\begin{equation}
R_x = \frac{2 x^{M} - x^{NR} - x^{ND} }{x^{NR} - x^{ND}}
\label{eq_isoratio} 
\end{equation}
where $x$ is an isospin-sensitive observable expected to be univocally related to the $N/Z$ of the systems under study, measured for the symmetric neutron-rich (NR) and neutron-deficient (ND) reactions, while the two mixed reactions (M) reach a neutron content in between these two references.
The evolution of $R_{x}$ towards $N/Z$ equilibration is thus followed as a function of a centrality parameter correlated to the dissipation of the collision.
By construction, $R_{x}=\pm1$, in the limit of fully nonequilibrated conditions (isospin transparency), while equilibration is generally signaled by both mixed reactions reaching the same value of the ratio. 
 
\subsection{Isospin diffusion in \textsc{amd-cc}}\label{subsec_model_study}

\begin{figure}[ht]
\includegraphics[scale=0.42]{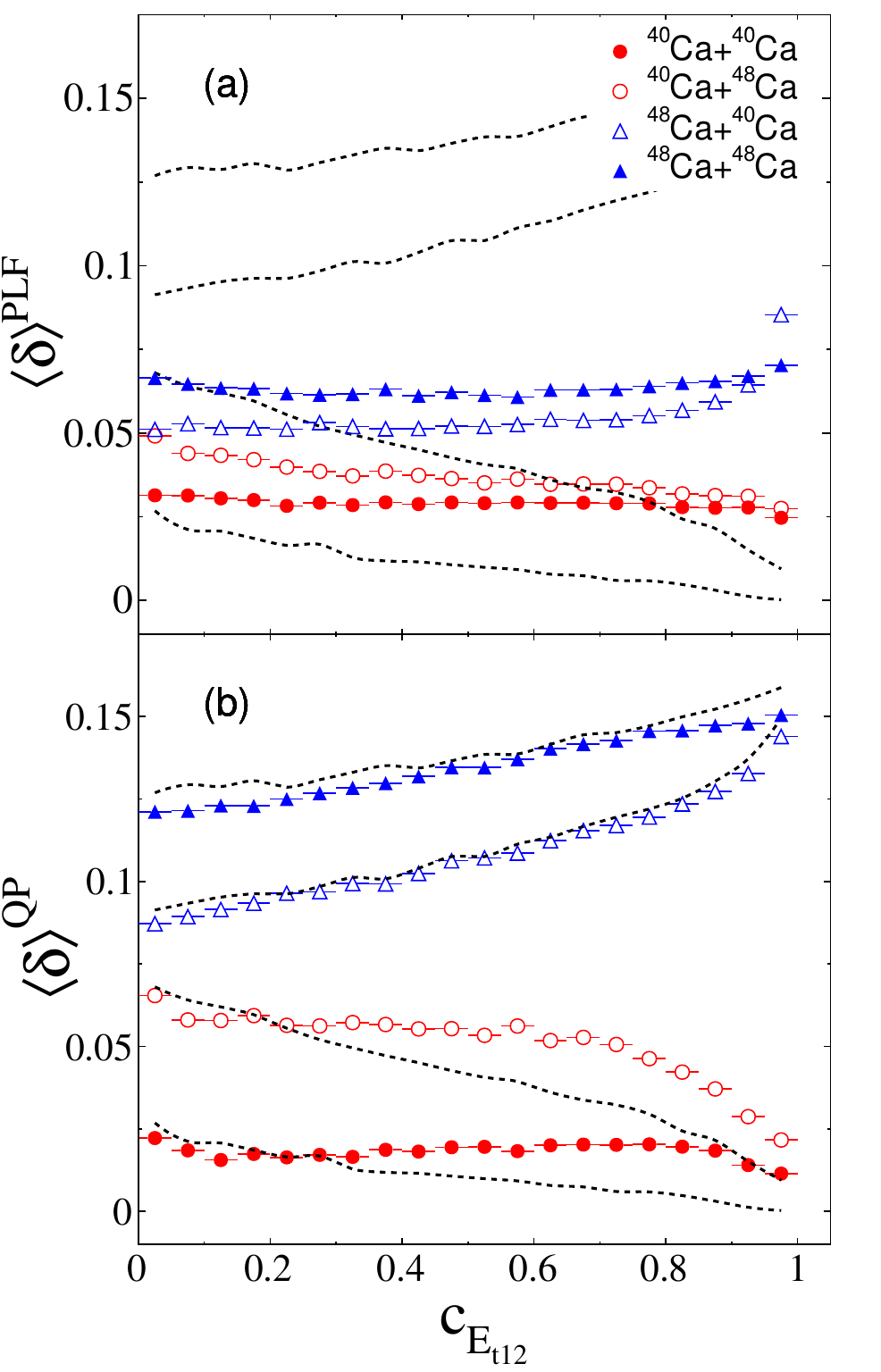}
\caption{Model calculations: distribution of the average asymmetry $\delta$ as a function of the centrality $c_{E_{t12}}$ for (a) the PLF and (b) the reconstructed QP, for \textsc{amd-cc} (stiff) filtered simulations (followed by \textsc{gemini++}).
The dashed lines correspond to the QP predicted by the filtered model before \textsc{gemini++}.
Statistical error bars are smaller than the symbols.}
\label{fig_delta_vs_cet12_mod}
\end{figure}

We present in Figs. \ref{fig_delta_vs_cet12_mod}(a) and \ref{fig_delta_vs_cet12_mod}(b) the evolution of the asymmetry $\delta=(N-Z)/A$ as a function of the centrality $c_{E_{t12}}$ from \textsc{amd-cc} (stiff) followed by \textsc{gemini++} for the PLF and the reconstructed QP, respectively.
We used the same sampling as the one used for the impact parameter estimation (20 bins of $5\%$ in $c_{E_{t12}}$). 
\begin{figure}[ht]
\centering
\includegraphics[scale=0.42]{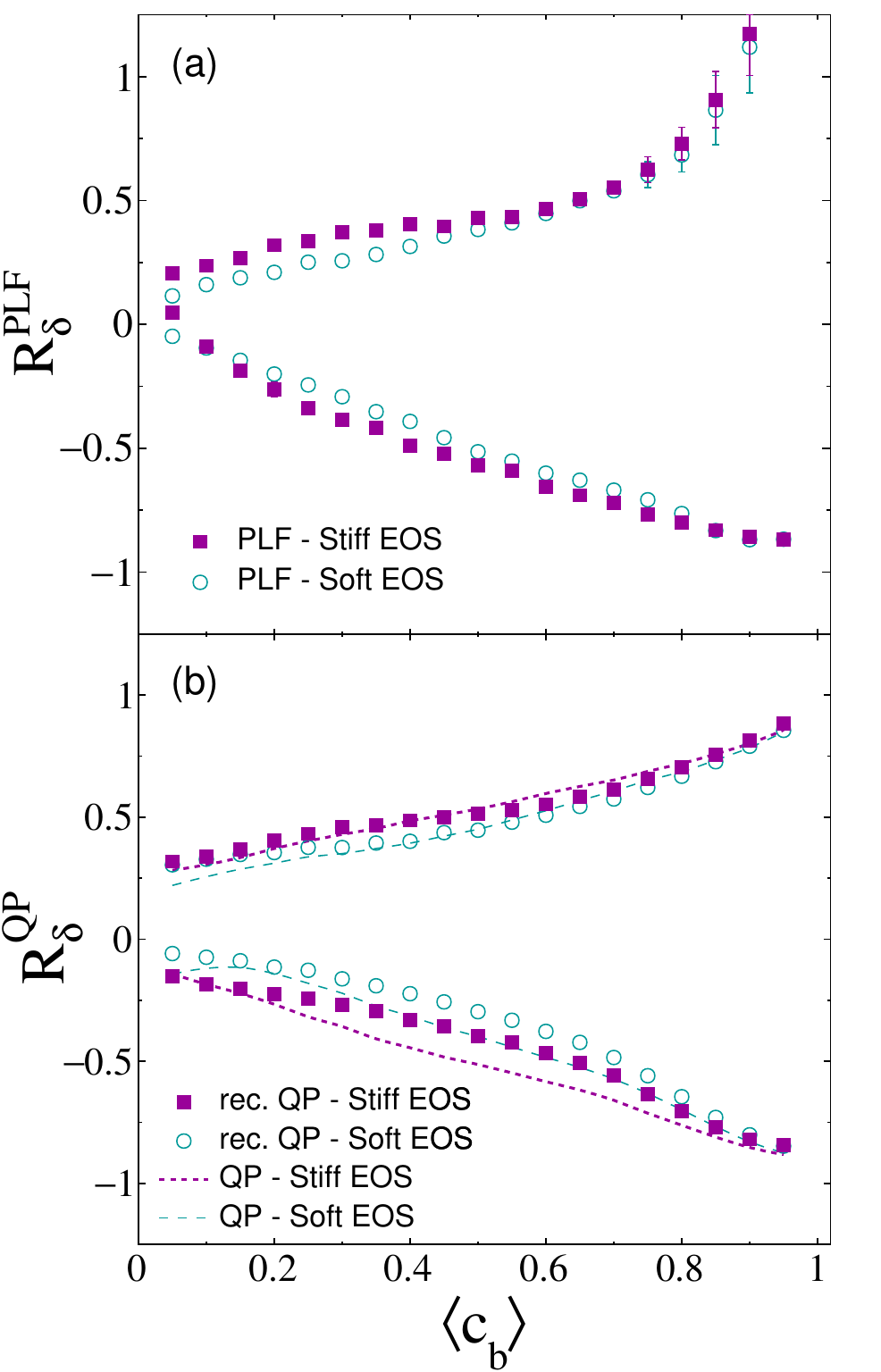}
\caption{Model calculations: isospin transport ratio as a function of the estimated average $\langle c_{b} \rangle$ for the PLF (a) and the reconstructed QP (b), respectively, from the \textsc{amd-cc} model with a stiff (full squares) and a soft (open circles) parametrization.
Results with the QP predicted by the model are superimposed in dashed lines.}
\label{fig_Rdelta_vs_cb_mod_stiff_soft}
\end{figure}
\begin{figure}[ht]
\centering
\includegraphics[scale=0.42]{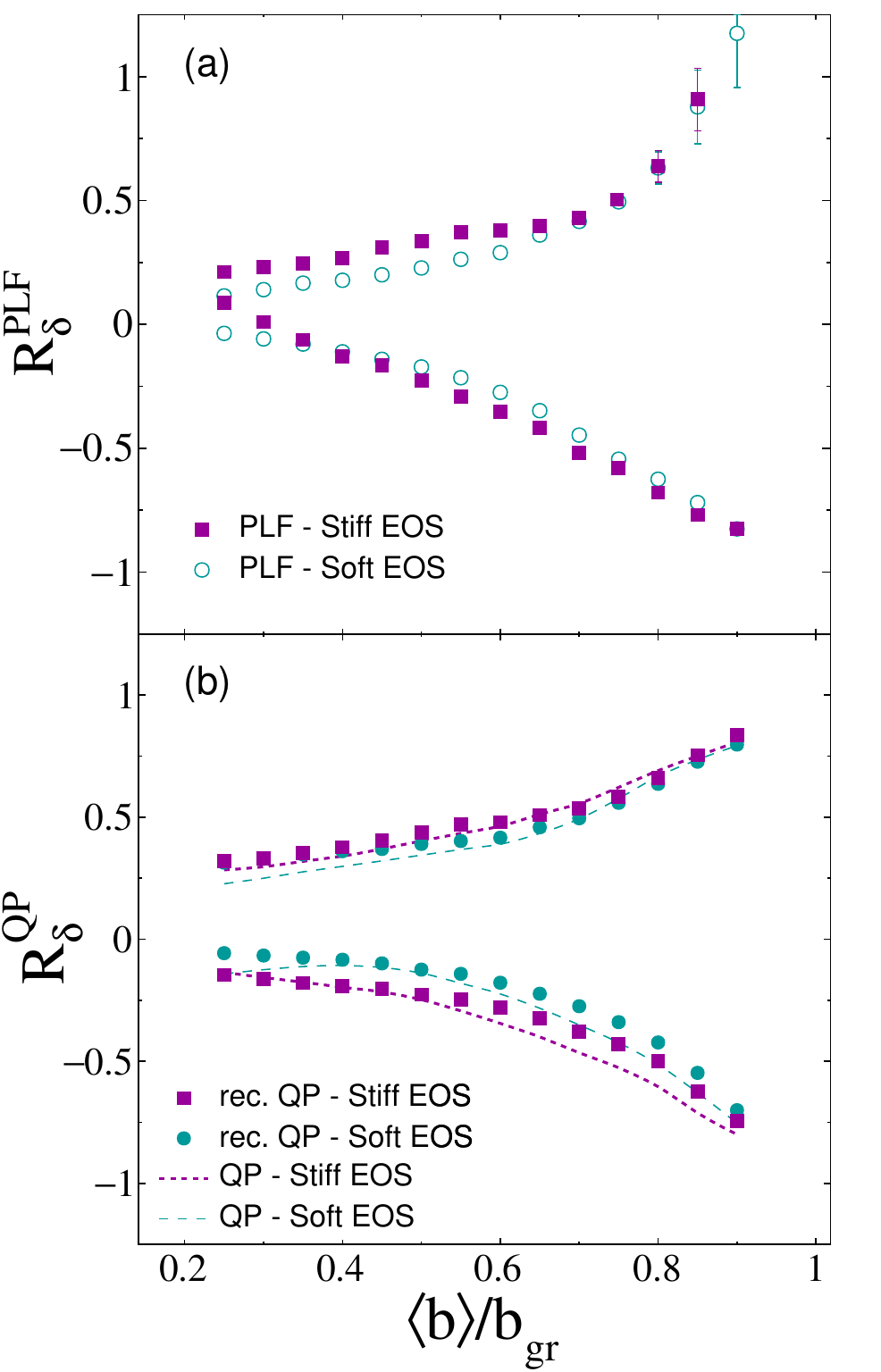}
\caption{Model calculations: isospin transport ratio as a function of the estimated average reduced impact parameter $\langle b \rangle / b_{gr}$ for the PLF (a) and the reconstructed QP (b), respectively, from the \textsc{amd-cc} model with a stiff (squares) and a soft (circles) parametrization.
Error bars are not shown when smaller than the symbols.
Results with the QP predicted by the model are shown in dashed lines.}
\label{fig_Rdelta_vs_bb_mod_stiff_soft}
\end{figure}

We first observe that both the PLF and the reconstructed QP exhibit the same hierarchy according to the neutron richness of the projectile and, to a lesser extent, of the target, as observed in the experimental data \cite{Fable_PRC_107_014604}.
The latter is also observed for the QP predicted by the model, represented in dashed line for each reaction.
Nonetheless, the model exhibits for all systems a significant difference between the asymmetry values obtained for the QP and its residue, resulting from the effect of sequential decays \cite{PhysRevC_107_054606}, that seems to be restored by the QP reconstruction method for $^{48}$Ca projectile reactions but not for $^{40}$Ca.

Using the estimated $\langle c_{b} \rangle$ and asymmetry $\delta$ for each sampling in $c_{E_{t12}}$, presented in Figs. \ref{fig_meancb_meanb_vs_cet12_mod}(a) and \ref{fig_delta_vs_cet12_mod} respectively, the corresponding isospin transport ratio can be computed from Eq. \ref{eq_isoratio} and plotted as a function of $\langle c_{b} \rangle$.
The results are presented in Figs. \ref{fig_Rdelta_vs_cb_mod_stiff_soft}(a) and \ref{fig_Rdelta_vs_cb_mod_stiff_soft}(b) for the PLF and the QP, respectively, for the stiff and soft parametrizations.
First, we observe converging values of the isospin transport ratio when moving from peripheral collisions (high $\langle c_{b} \rangle$ values) to more central collisions (low $\langle c_{b} \rangle$ values), indicating that an evolution towards isospin equilibration is predicted by the \textsc{amd-cc} model. 
Second, we notice that for the most central collisions probed ($\langle c_{b} \rangle < 0.1 $) the full equilibration condition is not reached. 
As observed in Fig. \ref{fig_meancb_meanb_vs_cet12_mod}(b), the average impact parameter expected to be probed in this region of centrality is about $\langle b \rangle \simeq 3 \pm 0.5$ fm, corresponding to semicentral collisions.
Third, the results indicate a weak sensitivity of the isospin transport ratio to the stiffness of the employed symmetry energy within the model, for both the PLF and the reconstructed QP.
Indeed, we notice that for the stiff asymmetry term (closed squares), $R_{\delta}$ exhibits systematically less isospin equilibration compared to the soft asymmetry term (open circles), for both the PLF and the reconstructed QP.
This is also observed from the QP predicted by the filtered model (not reconstructed), superimposed in thin and thick dashed lines for the soft and stiff parametrizations, respectively.
Such behavior has already been mentioned in \cite{Tsang2004:isospindiffusion} and \cite{Colonna_EPJA50} with Boltzmann-Uehling-Uhlenbeck (BUU) and stochastic mean-field approach (SMF) calculations, respectively.

In a similar way, the evolution of the isospin transport ratio as a function of the average reduced impact parameter, namely $\langle b \rangle/b_{gr}$, can be extracted from Figs. \ref{fig_meancb_meanb_vs_cet12_mod}(b) and \ref{fig_delta_vs_cet12_mod}.
The results are presented in Figs. \ref{fig_Rdelta_vs_bb_mod_stiff_soft}(a) and \ref{fig_Rdelta_vs_bb_mod_stiff_soft}(b) for the PLF and the QP, respectively, for the stiff and soft parametrizations.
We observe that the same conclusion as the one from Fig. \ref{fig_Rdelta_vs_cb_mod_stiff_soft} can be drawn, at least in the $\langle b \rangle/b_{gr}> 0.3$ region.
It is worth noting that the sensitivity to the stiffness of the employed symmetry energy remains weak, independently of the centrality or the QP reconstruction.

\subsection{Isospin diffusion in experimental data}\label{subsec_exp_study}

In this section, we apply the same protocol used with the \textsc{amd-cc} model to the INDRA-VAMOS experimental data.

Concerning the evaporated neutron estimation, the mean neutron multiplicities extracted from filtered \textsc{amd-cc} followed by \textsc{gemini++} calculations were used as a surrogate. 
More precisely, a constant scaling factor $\langle k \rangle = 0.8 $ is applied to the model neutron distribution in order to take into account the overstimation of light particles observed with \textsc{amd-cc} (see Sec. \ref{subsubsec_mult}). 
More details are given in Appendix \ref{app2_Mn}.

We present in Figs. \ref{fig_delta_vs_cet12_exp}(a) and \ref{fig_delta_vs_cet12_exp}(b) the evolution of the experimental asymmetry $\delta$ as a function of the centrality $c_{E_{t12}}$ (with a sampling of $5\%$) for the PLF and the reconstructed QP, respectively. 
\begin{figure}[ht]
\centering
\includegraphics[scale=0.42]{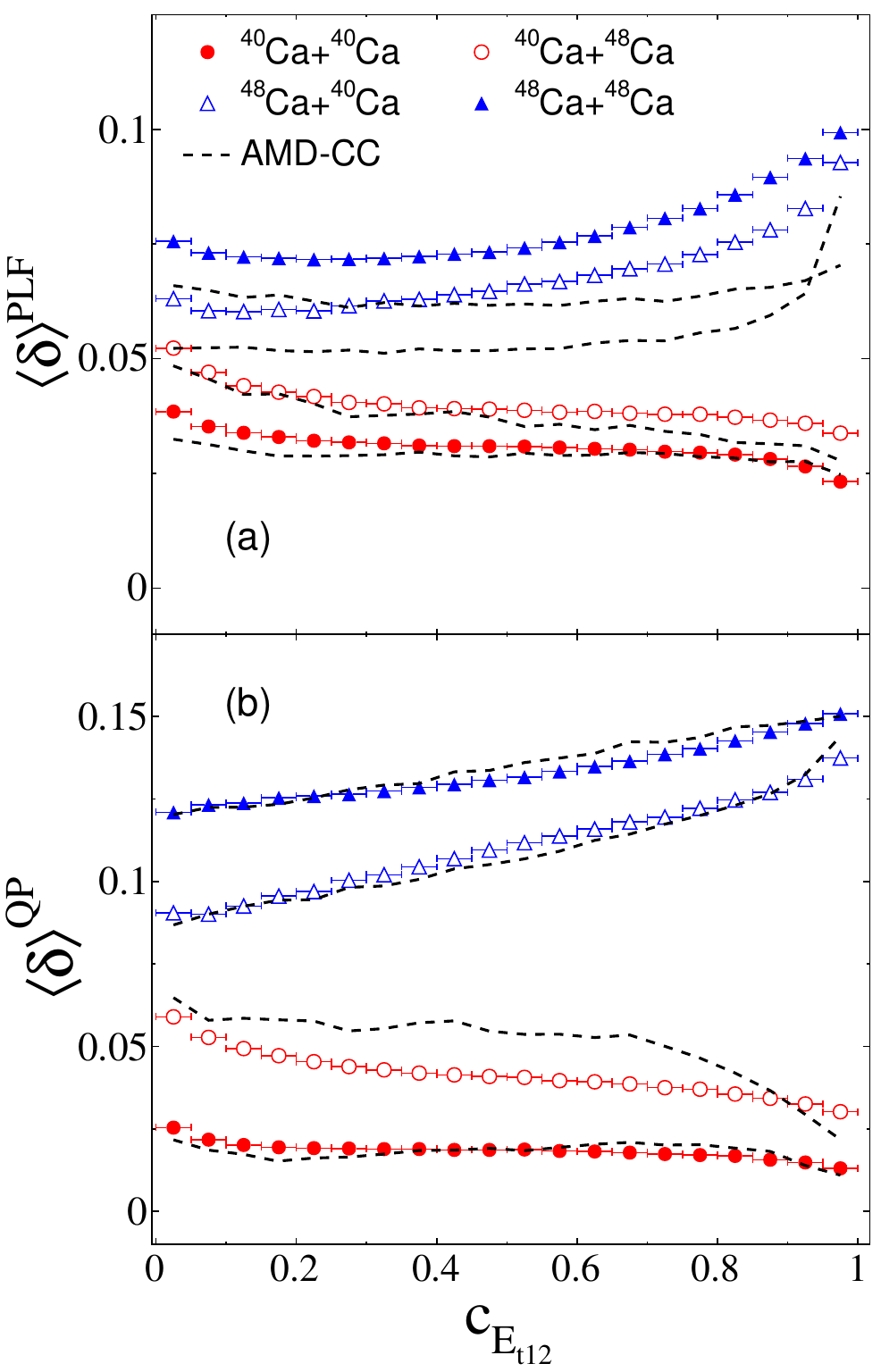}
\caption{Experimental data: distribution of the average asymmetry $\delta$ as a function of the centrality $c_{E_{t12}}$ for (a)
the PLF and (b) the reconstructed QP.
The dashed lines correspond to (a) the PLF and (b) the reconstructed QP obtained from \textsc{amd-cc} (stiff) filtered simulations (followed by \textsc{gemini++}). 
Statistical error bars are smaller than the symbols.}
\label{fig_delta_vs_cet12_exp}
\end{figure}
Concerning the PLF, we notice that the experimental data (symbols) exhibit systematically higher asymmetry values than \textsc{amd-cc} (dashed lines), while the same hierarchy is obtained. 
This effect is even more remarkable for the $^{48}$Ca projectile reactions, leading to the conclusion that \textsc{amd-cc} (followed by \textsc{gemini++}) fails to reproduce the experimental QP remnant neutron-enrichment over the full $c_{E_{t12}}$ domain.
Our results are consistent with the analysis of $^{40,48}$Ca+$^{40}$Ca reactions performed with four blocks of FAZIA, reported in \citep{CAMAIANI_PhysRevC_103_014605}.
Concerning the reconstructed QP, we observe that the applied reconstruction method leads to a better agreement between the experimental data and the model, except for the $^{40}$Ca$+^{48}$Ca asymmetric reaction.

%
\begin{figure}[ht]
\centering
\includegraphics[scale=0.42]{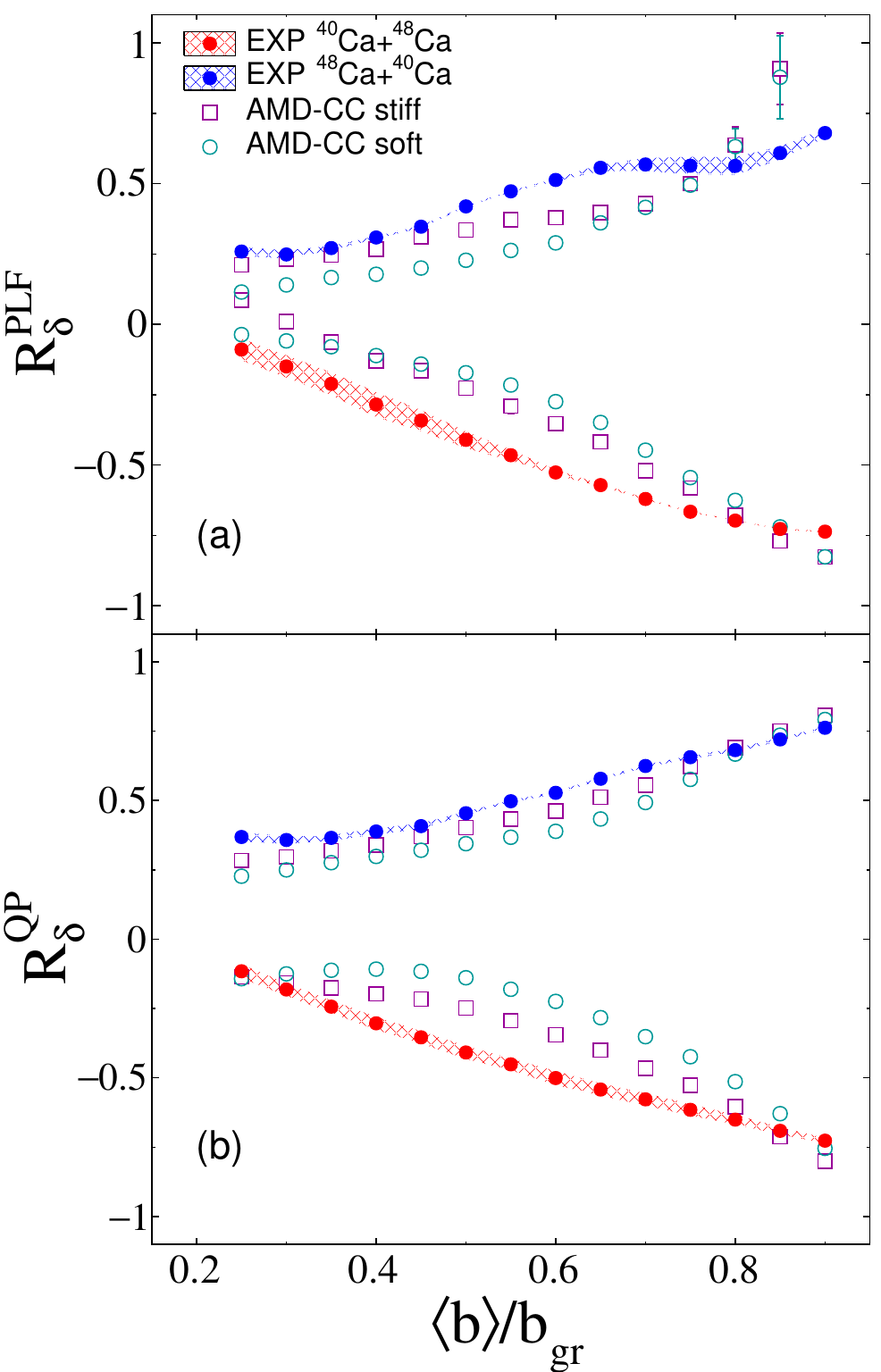}
\caption{Experimental isospin transport ratio computed from the asymmetry $\delta$ for the average reduced impact parameter $\langle b \rangle/b_{gr}$ estimated from the method described in Sec. \ref{sec_ip_recon}, for (a) the PLF and (b) the reconstructed QP. 
Open symbols represent the \textsc{amd-cc} filtered simulations (followed by \textsc{gemini++}) for a stiff (open squares) and a soft (open circles) symmetry energy.}
\label{fig_Rdelta_vs_bb_exp}
\end{figure}

Based on the results presented in Figs. \ref{fig_meancb_meanb_vs_cet12_exp}(b) and \ref{fig_delta_vs_cet12_exp}, the experimental isospin transport ratio was computed from Eq. \ref{eq_isoratio} as a function of the estimated reduced impact parameter $\langle b \rangle/b_{gr}$.
For more consistency, the scaling factor $k$ used for the evaporated neutron estimation was varied from $k = 0.7-0.9$, corresponding to a variation of one standard deviation of the average neutron multiplicity obtained for $\langle k \rangle = 0.8$, while the results presented thereafter are averaged over this domain in $k$.
The difference in average neutron multiplicity induced by the variation of $k$ is larger than the one induced by the stiffness of the symmetry energy.
Thus, this average also allows to remove the dependence of the average neutron multiplicity on the employed stiffness for the experimental QP reconstruction.

The results are presented in Figs. \ref{fig_Rdelta_vs_bb_exp}(a) and \ref{fig_Rdelta_vs_bb_exp}(b) for the experimental PLF and reconstructed QP (full circles), respectively.
It must also be noted that similar results can be drawn from the evolution of the experimental $R_{\delta}$ with $\langle c_{b} \rangle$. 
For both PLF and reconstructed QP, we first observe an evolution of the isospin transport ratio towards isospin equilibration, with decreasing values from about $R_{\delta} \simeq \pm 0.75$ towards $R_{\delta} \simeq -0.1$ and $R_{\delta} \simeq 0.3$ for the neutron-deficient and neutron-rich mixed reactions, respectively, when moving from the most peripheral ($\langle b \rangle/b_{gr} \simeq 0.9$) to the most central ($\langle b \rangle/b_{gr} \simeq 0.3$) measured collisions.
Furthermore, we remark a noticeable difference between the slopes of the ratios obtained from the two mixed reactions, leading to experimental values closer to the \textsc{amd-cc} stiff distributions for the $^{48}$Ca$+^{40}$Ca mixed system, while the PLF presents more equilibration than the QP at small impact parameters, consistent with the model (open symbols).
It is worth noting that a similar experimental trend is obtained for the PLF with the $^{48}$Ca+$^{40}$Ca mixed reaction in \citep{CAMAIANI_PhysRevC_103_014605}, where it was concluded that the observed discrepancy can be ascribed to an overestimated probability of nucleon transfer in \textsc{amd}.
In our understanding, the larger model-experiment discrepancy of $R_{\delta}$ in the $^{40}$Ca$+^{48}$Ca reaction originates from the significant difference in the evolution of the asymmetry $\delta$ with centrality $c_{E_{t12}}$, as observed in Fig. \ref{fig_delta_vs_cet12_exp}.   


\section{Summary and conclusions}\label{sec_conclusion}

In this work, we have presented an investigation of isospin diffusion in $^{40,48}$Ca$+^{40,48}$Ca reactions at 35 MeV/nucleon, by comparing INDRA-VAMOS experimental data to the output of the filtered \textsc{amd} model followed by \textsc{gemini++} as afterburner.

Two versions of \textsc{amd}, with and without cluster correlations, were employed in order to study the role of clustering in the global observables in the measured reactions.
Generally, both versions reproduce the experimental topology (charge and parallel velocity correlation).
Nonetheless, it was shown that the inclusion of clusters allows one to better reproduce the characteristics (velocity, mass and charge distributions) of the QP remnant detected in VAMOS (PLF).
Furthermore, comparisons of the average multiplicities have evidenced that clustering allows one to better reproduce the experimental multiplicities while it is mandatory to reproduce the LCP transverse kinetic energy distributions.
It is also important to note that, aside from clustering, the treatment of dissipation (collision term), nuclear stopping and excitation energy within the models plays an important role in reproducing the experimental data.   
We believe that reproducing the global observables outlined in this study sets a fundamental standard for the trustworthiness of dynamic models. 
This standard ensures that, when comparing models to experimental data, the comparison becomes meaningful and enhances the precision in constraining the equation of state of nuclear matter.

We have applied a method for impact parameter reconstruction specifically adapted to the Fermi energy domain, allowing us to infer information on the centrality (and the impact parameter) from the inclusive distribution of an experimentally measured observable.
The method was adapted to INDRA-VAMOS, using the inclusive distribution of total transverse kinetic energy, and tested within the \textsc{amd-cc} model, proving its relevance over the whole impact parameter domain.        

The isospin diffusion phenomenon was investigated by means of the isospin transport ratio computed from the asymmetry $\delta = (N-Z)/A$ of both PLF and reconstructed QP.
For the first time, we have highlighted that the employed impact parameter reconstruction method allows a more direct comparison to the experimental data for the study of isospin diffusion.
Furthermore, our results show that the weak sensitivity of the isospin transport ratio to the stiffness of the employed symmetry energy, expected for the QP predicted by the model (before sequential decays), holds for both the PLF and the reconstructed QP.
Nonetheless, a noticeable disagreement is observed in the centrality dependence of the ratios between the experimental data and the model, more specifically for the $^{40}$Ca$+^{48}$Ca mixed system.
Such difference can be anticipated from the individual evolution of $\delta$ with centrality obtained from the PLF, that the \textsc{amd} model followed by \textsc{gemini++} fails to reproduce for all reactions.
In our understanding, this significant discrepancy originates from the overestimation of the excitation energy of the QP predicted by both \textsc{amd} models and inputted in the afterburner.
Consequently, for the neutron rich $^{48}$Ca projectile reactions the evolution of the average neutron excess is mostly driven by the evaporative attractor line, while the reconstruction of the QP indicates a better agreement between the model and the experimental data, as compared to the $^{40}$Ca projectile reactions.
We believe that the issue of the overestimation of the excitation energy per nucleon in \textsc{amd} is critical to improve the model predictions on isospin transport.

The results presented in this work constitute a further step to improve the comparison protocol employed in the study of isospin transport and of the nuclear EoS. 
The subsequent phase involves implementing the suggested protocol across a variety of dynamical models, including both QMD-like and BUU-like formalisms, for which we can anticipate that variations in the mean-field implementation will result in distinct behaviors in the isospin transport ratio.
This work is currently in progress.
 		 
\begin{acknowledgments}
The authors would like to thank:
The technical staff of GANIL for their continued support for performing the experiments;
The CNRS/IN2P3 Computing Center (Lyon, France) for providing data-processing resources needed for this work.
This work was partly supported by the IN2P3-GSI agreement (Grant No. 03-45); 
The National Research Foundation of Korea (NRF) (Grant No. 2018R1A5A1025563);
IBS (Grant No. IBS-R031-D1) in Korea.
Q. F. gratefully acknowledges the support from CNRS-IN2P3.
\end{acknowledgments}

\appendix

\section{Impact parameter reconstruction}\label{app1_ip}

\subsection{Method}\label{subsec_ip_method}

By definition, the inclusive distribution of an observable $X$, $P(X)$, resulting from all measured collisions with an unknown impact parameter distribution $P(b)$, is related to the conditional probability distribution of $X$ at fixed $b$, $P(X|b)$, such as
\begin{equation}
P(X) \equiv \int_{0}^{+\infty} P(b)P(X|b) db = \int_{0}^{1} P(X|c_{b}) dc_{b}
\label{eq_PX_inclusive}
\end{equation}

The right-hand side of Eq. \ref{eq_PX_inclusive} is obtained by introducing the centrality $c_{b}$, defined as the cumulative distribution function of $P(b)$: 
\begin{equation}
c_{b} \equiv \int_{0}^{b} P(b') db'
\label{eq_cb}
\end{equation}

with $P(c_{b})=1$ $\forall c_{b}$. 
As described in \cite{PhysRevC_98_024902, PhysRevC_104_034609}, Eq. \ref{eq_PX_inclusive} can be used to determine $P(X|c_{b})$ by fitting the experimentally measured inclusive distribution $P(X)$, with a suitable probability density function (p.d.f.) which encodes both the centrality dependence of the mean value $\overline{X}(c_{b})$ and the fluctuation of $X$ about this mean value.

Once $P(X|c_{b})$ is obtained by fitting the experimental $P(X)$ distribution, the centrality distribution of any experimental generic sample $\mathbb{S}$, $P(c_{b}|\mathbb{S})$, can be deduced from the Bayes's theorem:
\begin{equation}
P(c_{b}|\mathbb{S}) = \frac{\int P(X|c_{b}) \frac{P(X|\mathbb{S})}{P(X)} dX}{\int P(X|\mathbb{S}) dX}
\label{eq_PCb_for_S}
\end{equation}

where $P(X|\mathbb{S})$ is an histogram of $X$ filled with the events of the sample while the integrals are performed over the full domain of $X$.

Finally, the absolute impact parameter distribution can be deduced from the previously calculated centrality distribution, as
\begin{equation}
P(b|\mathbb{S}) = P(b) P(c_{b}(b)|\mathbb{S})
\label{eq_Pb_for_S}
\end{equation}

The method described above is model independent and allows one to take into account the fluctuations inherent to the relationship between any experimental observable and the impact parameter.
It should nonetheless be noted that it is necessary to assume a specific form for $P(b)$ in order to use Eq. \ref{eq_PCb_for_S} and calculate $c_{b}(b)$, \textit{i.e.}, the relation between $c_{b}$ and $b$.

\subsection{Implementation for INDRA-VAMOS data}\label{subsec_ip_approx}

In the present analysis the impact parameter distributions are reconstructed by using the inclusive distributions in total transverse energy of the LCP, namely $P(E_{t12})$, where $E_{t12}$ is defined in Eq. \ref{eq_et12}. 
It is worth noting that the PLF properties are largely independent of this quantity, avoiding possible trivial bias due to autocorrelation with the event sorting for the study of the isospin transport ratio \cite{Rami_PhysRevLett_84_1120}.
Also, as can be seen in Fig. \ref{fig_Et12}, this quantity is relatively well reproduced by the \textsc{amd-cc} filtered model, while we have verified that it presents a monotonic relationship with centrality $c_{b}$ within the model.

Second, the data are sampled according to the experimental centrality $c_{E_{t12}}$, defined as the complementary cumulative distribution function of the $P(E_{t12})$ distribution:
\begin{equation}
c_{E_{t12}}\equiv\int_{E_{t12}}^{+\infty}P(\tilde{E}_{t12})\:\mathrm{d}\tilde{E}_{t12}\label{eq_cent}
\end{equation}
By construction, $c_{E_{t12}}$ decreases from 1 to 0 as $E_{t12}$ goes from its minimum ($\approx 0$) to its (system-dependent) maximum value. 
Therefore, large ($\approx 1$) $c_{E_{t12}}$ values are associated with the most peripheral collisions while smaller values ($c_{E_{t12}}\rightarrow0$) indicate smaller average impact parameters. 
This choice is based on our previous work and aims to remove the system dependence of the $E_{t12}$ distributions in the data sampling \cite{Fable_PRC_107_014604}.
Indeed, since bins of fixed width of $c_{E_{t12}}$ contain the same number of events, we expect them to have the same statistical significance, whatever the centrality.
Based on the aforementioned considerations, a fixed step of $5\%$ experimental centrality $c_{E_{t12}}$ have been used for the data sampling, for a total of 20 bin samples.

Third, we have adopted the same specific implementation as Frankland \textit{et~al.} for the fluctuation kernel and the centrality dependence of $\overline{E}_{t12}$.
Concerning the fluctuation kernel, since $E_{t12}$ is positive the gamma distribution is an ideal choice \cite{PhysRevC_98_024902}.
Accordingly, the p.d.f. used to fit the inclusive distribution reads
\begin{equation}
P(X|c_{b}) = \frac{1}{\Gamma(k)\theta^k}X^{k-1}e^{-X/\theta}
\label{eq_gamma}
\end{equation}

where $k$ and $\theta$ are two positive parameters related to the mean value $\overline{X}$ and its variance, respectively.
As pointed out in \cite{PhysRevC_98_024902}, these parameters generally depend on the centrality, while the fit to the inclusive distribution $P(X)$ using Eq. \ref{eq_gamma} to extract $k(c_{b})$ and $\theta(c_{b})$ is underconstrained.
In the framework of the gamma-distribution of Eq. \ref{eq_gamma}, the centrality dependence of $k$ must be parametrized, and it is generally assumed that the variance parameter $\theta$ is constant for all centralities. 
Therefore, similarly to \cite{PhysRevC_104_034609}, in this work $k$ is parametrized with a monotonically decreasing function of centrality, while $\theta$ is a free parameter of the fit constrained by the tail of $P(X)$.
In summary, five free parameters $(\alpha, \gamma, \theta, X_{min}, X_{max})$ are required for the fit; they are explicitly defined in \cite{PhysRevC_104_034609}.

Finally, we would like to stress that if the model calculations were run with a geometric impact parameter distribution, the filtered impact parameter distribution $P(b)$ cannot be well-fitted by the usual approximation observed for most of the INDRA dataset (see Eq.12 in \cite{PhysRevC_104_034609}).
This can be seen in Fig. \ref{fig_b}, where we clearly observe the effect of the VAMOS angular cuts that favor the detection of semiperipheral and peripheral events, even for \textsc{amd-cc}, in the $b/b_{gr}=0.6-0.9$ region.
For this reason, and at the cost of losing the model independence for $P(X|\mathbb{S})$, we have used the inclusive impact parameter distribution directly obtained from the filtered model as a surrogate for $P(b)$ in Eq. \ref{eq_Pb_for_S}.
Therefore, for more consistency the results on isospin diffusion will be presented using both Eqs. \ref{eq_PCb_for_S} and \ref{eq_Pb_for_S}, keeping in mind that the former is model independent. 

\section{Quasiprojectile reconstruction}\label{app2_Mn}

Similarly to our previous works, the relative velocities between the reactions products measured (filtered) in INDRA and (i) the PLF detected (filtered) in VAMOS and (ii) the fragment with the largest identified charge at backward angles (supposed to be the TLF) are calculated.
Numerically, cuts on the associated relative velocities, respectively $V_{rel,PLF}$ and $V_{rel,TLF}$, were applied in order to include fragments whose velocities verify $\frac{V_{rel,TLF}}{V_{rel,PLF}}>1.35$ for $Z=1$ and $\frac{V_{rel,TLF}}{V_{rel,PLF}}>1.75$ for $Z \geq 2$.
The values of the cutoff thresholds were estimated from both \textsc{amd-nc} and \textsc{amd-cc} filtered simulations, to optimize the contribution of the actual QP to the reconstruction.
Interestingly, the values of the cuts are similar for both versions of \textsc{amd}, while their effect remains consistent for the models and the experiment.  

The quasiprojectile atomic number $Z_{QP}$ reads
\begin{equation}\label{eq_zqp}
Z_{QP} = Z_{PLF} + \sum_{i}^{M_{I}} Z_{i}
\end{equation}

where $Z_{PLF}$ is the charge of the fragment measured (filtered) in VAMOS and the sum runs over the charge $Z_{i}$ of the $M_{I}$ selected particles. 

Similarly, the quasiprojectile mass number (without evaporated neutron contribution) $\tilde{A}_{QP}$ reads
\begin{equation}\label{eq_aqp}
\tilde{A}_{QP} = A_{PLF} + \sum_{i}^{M_{I}} A_{i}
\end{equation}

where $A_{PLF}$ is the mass of the fragment in VAMOS and the sum runs over the mass $A_{i}$ of the $M_{I}$ selected particles. 

%
\begin{figure}[h]
\centering
\includegraphics[scale=0.45]{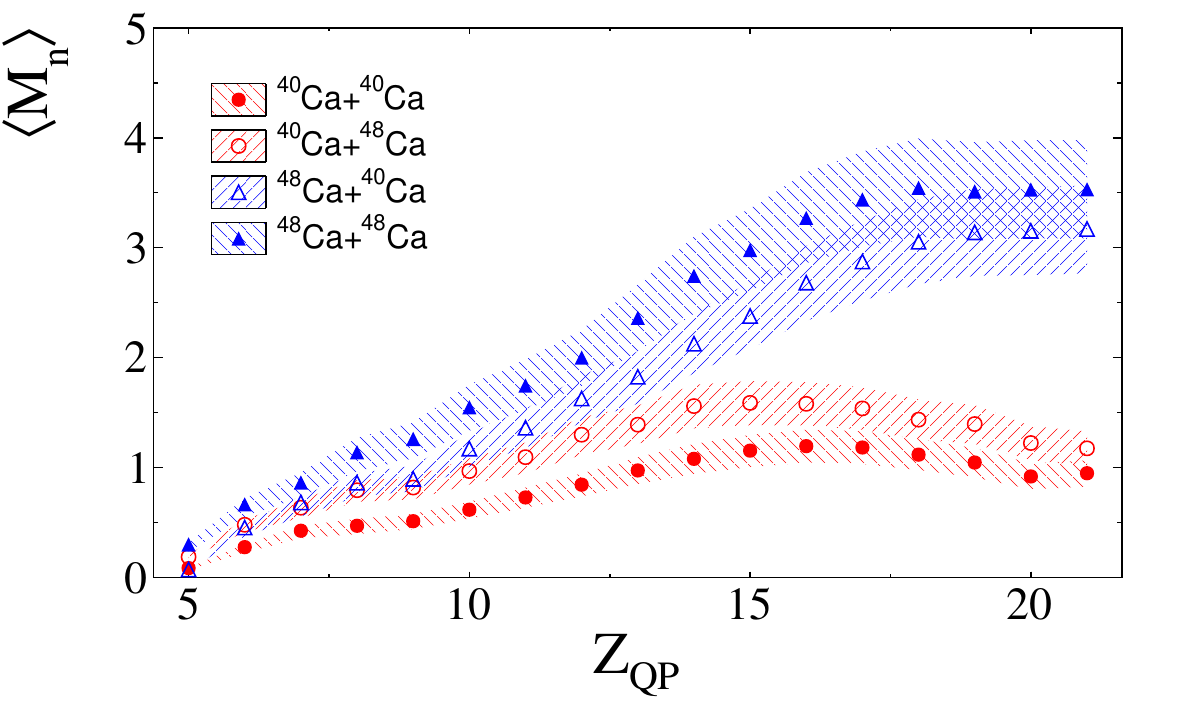}
\caption{Average neutron multiplicities distributions used for the mass reconstruction of the experimental quasiprojectile as a function of its charge.
The symbols correspond to the values obtained for $\langle k \rangle = 0.8$ while the dashed areas represent a variation of $k=0.7-0.9$ (roughly one standard deviation).}
\label{fig_mn_exp}
\end{figure}

As the neutrons were not measured in the INDRA-VAMOS experiment, the distributions of the neutrons evaporated by the reconstructed QP from the \textsc{amd} filtered model calculations were used as a substitute.
More precisely, for each event with reconstructed charge $Z_{QP}$ and mass without the neutron $\tilde{A}_{QP}$, the evaporated neutron multiplicity was estimated from a random number generator following the filtered model neutron multiplicity distribution (histogram).

The estimated evaporated neutron multiplicity is thus the following:
\begin{equation}\label{eq_mn}
M_n (Z_{QP}, \tilde{A}_{QP}) = \lceil M_n^{rdm}(Z_{QP}, \tilde{A}_{QP}) \cdot k \rceil
\end{equation}
where $M_n^{rdm}$ is the random neutron multiplicity extracted from the model histogram, $k$ is a correction factor and $\lceil \rceil$ is the ceiling function.

The scaling factor $k$ is intended to correct from the fact that both \textsc{amd} models tends to overestimate the proton multiplicities compared to the experiment, as observed in Sec. \ref{subsubsec_mult}.

Thus, assuming that the experimental and filtered model average neutron-to-proton multiplicity ratios are equivalent, we have

\begin{equation}\label{eq_kratio}
\langle M_n \rangle^{exp} = \langle M_n \rangle^{mod} \cdot \frac{\langle M_p \rangle ^{exp}}{\langle M_p \rangle ^{mod}} = \langle M_n \rangle^{mod} \cdot k 
\end{equation}
where $\langle M_{n,p} \rangle^{exp,mod}$ are the neutron and proton average multiplicities of the experiment and model, respectively.

Therefore, $k=1$ when applying the evaporated neutron correction within the model, while a constant value of $\langle k \rangle = 0.8$ is applied in the case of the experiment, which corresponds to the average scaling factor over all systems and charges.
The resulting experimental average neutron distributions are presented in Fig. \ref{fig_mn_exp}, for the four systems under study.
In this figure, the symbols correspond to the values obtained for $\langle k \rangle = 0.8$, while the dashed areas represent a variation from $k=0.7-0.9$, corresponding to a variation of one standard deviation.

\bibliographystyle{apsrev}
\addcontentsline{toc}{section}{\refname}
\bibliography{extracted.bib}

\begin{thebibliography}{51}
\expandafter\ifx\csname natexlab\endcsname\relax\def\natexlab#1{#1}\fi
\expandafter\ifx\csname bibnamefont\endcsname\relax
  \def\bibnamefont#1{#1}\fi
\expandafter\ifx\csname bibfnamefont\endcsname\relax
  \def\bibfnamefont#1{#1}\fi
\expandafter\ifx\csname citenamefont\endcsname\relax
  \def\citenamefont#1{#1}\fi
\expandafter\ifx\csname url\endcsname\relax
  \def\url#1{\texttt{#1}}\fi
\expandafter\ifx\csname urlprefix\endcsname\relax\def\urlprefix{URL }\fi
\providecommand{\bibinfo}[2]{#2}
\providecommand{\eprint}[2][]{\url{#2}}

\bibitem[{\citenamefont{Lattimer and Prakash}(2016)}]{LATTIMER2016127}
\bibinfo{author}{\bibfnamefont{J.~M.} \bibnamefont{Lattimer}} \bibnamefont{and}
  \bibinfo{author}{\bibfnamefont{M.}~\bibnamefont{Prakash}},
  \bibinfo{journal}{Physics Reports} \textbf{\bibinfo{volume}{621}},
  \bibinfo{pages}{127} (\bibinfo{year}{2016}), ISSN \bibinfo{issn}{0370-1573},
  \bibinfo{note}{memorial Volume in Honor of Gerald E. Brown}.

\bibitem[{\citenamefont{Toro et~al.}(2010)\citenamefont{Toro, Baran, Colonna,
  and Greco}}]{Di_Toro_2010}
\bibinfo{author}{\bibfnamefont{M.~D.} \bibnamefont{Toro}},
  \bibinfo{author}{\bibfnamefont{V.}~\bibnamefont{Baran}},
  \bibinfo{author}{\bibfnamefont{M.}~\bibnamefont{Colonna}}, \bibnamefont{and}
  \bibinfo{author}{\bibfnamefont{V.}~\bibnamefont{Greco}},
  \bibinfo{journal}{Journal of Physics G: Nuclear and Particle Physics}
  \textbf{\bibinfo{volume}{37}}, \bibinfo{pages}{083101}
  (\bibinfo{year}{2010}),
  \urlprefix\url{https://dx.doi.org/10.1088/0954-3899/37/8/083101}.

\bibitem[{\citenamefont{Tsang et~al.}(2012)\citenamefont{Tsang, Stone, Camera,
  Danielewicz, Gandolfi, Hebeler, Horowitz, Lee, Lynch, Kohley
  et~al.}}]{PhysRevC_86_015803}
\bibinfo{author}{\bibfnamefont{M.~B.} \bibnamefont{Tsang}},
  \bibinfo{author}{\bibfnamefont{J.~R.} \bibnamefont{Stone}},
  \bibinfo{author}{\bibfnamefont{F.}~\bibnamefont{Camera}},
  \bibinfo{author}{\bibfnamefont{P.}~\bibnamefont{Danielewicz}},
  \bibinfo{author}{\bibfnamefont{S.}~\bibnamefont{Gandolfi}},
  \bibinfo{author}{\bibfnamefont{K.}~\bibnamefont{Hebeler}},
  \bibinfo{author}{\bibfnamefont{C.~J.} \bibnamefont{Horowitz}},
  \bibinfo{author}{\bibfnamefont{J.}~\bibnamefont{Lee}},
  \bibinfo{author}{\bibfnamefont{W.~G.} \bibnamefont{Lynch}},
  \bibinfo{author}{\bibfnamefont{Z.}~\bibnamefont{Kohley}},
  \bibnamefont{et~al.}, \bibinfo{journal}{Phys. Rev. C}
  \textbf{\bibinfo{volume}{86}}, \bibinfo{pages}{015803}
  (\bibinfo{year}{2012}),
  \urlprefix\url{https://link.aps.org/doi/10.1103/PhysRevC.86.015803}.

\bibitem[{\citenamefont{Lattimer and Lim}(2013)}]{Lattimer_2013}
\bibinfo{author}{\bibfnamefont{J.~M.} \bibnamefont{Lattimer}} \bibnamefont{and}
  \bibinfo{author}{\bibfnamefont{Y.}~\bibnamefont{Lim}}, \bibinfo{journal}{The
  Astrophysical Journal} \textbf{\bibinfo{volume}{771}}, \bibinfo{pages}{51}
  (\bibinfo{year}{2013}),
  \urlprefix\url{https://dx.doi.org/10.1088/0004-637X/771/1/51}.

\bibitem[{\citenamefont{Huth et~al.}(2022)\citenamefont{Huth, Pang, Tews,
  Dietrich, Le~F{\`e}vre, Schwenk, Trautmann, Agarwal, Bulla, Coughlin
  et~al.}}]{Huth2022}
\bibinfo{author}{\bibfnamefont{S.}~\bibnamefont{Huth}},
  \bibinfo{author}{\bibfnamefont{P.~T.~H.} \bibnamefont{Pang}},
  \bibinfo{author}{\bibfnamefont{I.}~\bibnamefont{Tews}},
  \bibinfo{author}{\bibfnamefont{T.}~\bibnamefont{Dietrich}},
  \bibinfo{author}{\bibfnamefont{A.}~\bibnamefont{Le~F{\`e}vre}},
  \bibinfo{author}{\bibfnamefont{A.}~\bibnamefont{Schwenk}},
  \bibinfo{author}{\bibfnamefont{W.}~\bibnamefont{Trautmann}},
  \bibinfo{author}{\bibfnamefont{K.}~\bibnamefont{Agarwal}},
  \bibinfo{author}{\bibfnamefont{M.}~\bibnamefont{Bulla}},
  \bibinfo{author}{\bibfnamefont{M.~W.} \bibnamefont{Coughlin}},
  \bibnamefont{et~al.}, \bibinfo{journal}{Nature}
  \textbf{\bibinfo{volume}{606}}, \bibinfo{pages}{276} (\bibinfo{year}{2022}),
  ISSN \bibinfo{issn}{1476-4687}.

\bibitem[{\citenamefont{Lynch and Tsang}(2022)}]{LYNCH2022137098}
\bibinfo{author}{\bibfnamefont{W.}~\bibnamefont{Lynch}} \bibnamefont{and}
  \bibinfo{author}{\bibfnamefont{M.}~\bibnamefont{Tsang}},
  \bibinfo{journal}{Physics Letters B} \textbf{\bibinfo{volume}{830}},
  \bibinfo{pages}{137098} (\bibinfo{year}{2022}), ISSN
  \bibinfo{issn}{0370-2693},
  \urlprefix\url{https://www.sciencedirect.com/science/article/pii/S0370269322002325}.

\bibitem[{\citenamefont{Tsang et~al.}(2024)\citenamefont{Tsang, Tsang, Lynch,
  Kumar, and Horowitz}}]{Tsang2024}
\bibinfo{author}{\bibfnamefont{C.~Y.} \bibnamefont{Tsang}},
  \bibinfo{author}{\bibfnamefont{M.~B.} \bibnamefont{Tsang}},
  \bibinfo{author}{\bibfnamefont{W.~G.} \bibnamefont{Lynch}},
  \bibinfo{author}{\bibfnamefont{R.}~\bibnamefont{Kumar}}, \bibnamefont{and}
  \bibinfo{author}{\bibfnamefont{C.~J.} \bibnamefont{Horowitz}},
  \bibinfo{journal}{Nature Astronomy}  (\bibinfo{year}{2024}), ISSN
  \bibinfo{issn}{2397-3366},
  \urlprefix\url{https://doi.org/10.1038/s41550-023-02161-z}.

\bibitem[{\citenamefont{Russotto et~al.}(2011)\citenamefont{Russotto, Wu,
  Zoric, Chartier, Leifels, Lemmon, Li, \L{}ukasik, Pagano, Paw\l{}owski
  et~al.}}]{RUSSOTTO2011471}
\bibinfo{author}{\bibfnamefont{P.}~\bibnamefont{Russotto}},
  \bibinfo{author}{\bibfnamefont{P.}~\bibnamefont{Wu}},
  \bibinfo{author}{\bibfnamefont{M.}~\bibnamefont{Zoric}},
  \bibinfo{author}{\bibfnamefont{M.}~\bibnamefont{Chartier}},
  \bibinfo{author}{\bibfnamefont{Y.}~\bibnamefont{Leifels}},
  \bibinfo{author}{\bibfnamefont{R.}~\bibnamefont{Lemmon}},
  \bibinfo{author}{\bibfnamefont{Q.}~\bibnamefont{Li}},
  \bibinfo{author}{\bibfnamefont{J.}~\bibnamefont{\L{}ukasik}},
  \bibinfo{author}{\bibfnamefont{A.}~\bibnamefont{Pagano}},
  \bibinfo{author}{\bibfnamefont{P.}~\bibnamefont{Paw\l{}owski}},
  \bibnamefont{et~al.}, \bibinfo{journal}{Physics Letters B}
  \textbf{\bibinfo{volume}{697}}, \bibinfo{pages}{471} (\bibinfo{year}{2011}),
  ISSN \bibinfo{issn}{0370-2693},
  \urlprefix\url{https://www.sciencedirect.com/science/article/pii/S037026931100178X}.

\bibitem[{\citenamefont{Abbott et~al.}(2017)\citenamefont{Abbott, Abbott,
  Abbott, Acernese, Ackley, Adams, Adams, Addesso, Adhikari, Adya
  et~al.}}]{Abbott_PhysRevLett_119_161101}
\bibinfo{author}{\bibfnamefont{B.~P.} \bibnamefont{Abbott}},
  \bibinfo{author}{\bibfnamefont{R.}~\bibnamefont{Abbott}},
  \bibinfo{author}{\bibfnamefont{T.~D.} \bibnamefont{Abbott}},
  \bibinfo{author}{\bibfnamefont{F.}~\bibnamefont{Acernese}},
  \bibinfo{author}{\bibfnamefont{K.}~\bibnamefont{Ackley}},
  \bibinfo{author}{\bibfnamefont{C.}~\bibnamefont{Adams}},
  \bibinfo{author}{\bibfnamefont{T.}~\bibnamefont{Adams}},
  \bibinfo{author}{\bibfnamefont{P.}~\bibnamefont{Addesso}},
  \bibinfo{author}{\bibfnamefont{R.~X.} \bibnamefont{Adhikari}},
  \bibinfo{author}{\bibfnamefont{V.~B.} \bibnamefont{Adya}},
  \bibnamefont{et~al.}, \bibinfo{journal}{Phys. Rev. Lett.}
  \textbf{\bibinfo{volume}{119}}, \bibinfo{pages}{161101}
  (\bibinfo{year}{2017}).

\bibitem[{\citenamefont{Typel}(2013)}]{Typel_2013}
\bibinfo{author}{\bibfnamefont{S.}~\bibnamefont{Typel}},
  \bibinfo{journal}{Journal of Physics: Conference Series}
  \textbf{\bibinfo{volume}{420}}, \bibinfo{pages}{012078}
  (\bibinfo{year}{2013}),
  \urlprefix\url{https://dx.doi.org/10.1088/1742-6596/420/1/012078}.

\bibitem[{\citenamefont{Brown}(2013)}]{PhysRevLett_111_232502}
\bibinfo{author}{\bibfnamefont{B.~A.} \bibnamefont{Brown}},
  \bibinfo{journal}{Phys. Rev. Lett.} \textbf{\bibinfo{volume}{111}},
  \bibinfo{pages}{232502} (\bibinfo{year}{2013}),
  \urlprefix\url{https://link.aps.org/doi/10.1103/PhysRevLett.111.232502}.

\bibitem[{\citenamefont{Danielewicz and Lee}(2014)}]{DANIELEWICZ20141}
\bibinfo{author}{\bibfnamefont{P.}~\bibnamefont{Danielewicz}} \bibnamefont{and}
  \bibinfo{author}{\bibfnamefont{J.}~\bibnamefont{Lee}},
  \bibinfo{journal}{Nuclear Physics A} \textbf{\bibinfo{volume}{922}},
  \bibinfo{pages}{1} (\bibinfo{year}{2014}), ISSN \bibinfo{issn}{0375-9474},
  \urlprefix\url{https://www.sciencedirect.com/science/article/pii/S0375947413007872}.

\bibitem[{\citenamefont{Russotto et~al.}(2016)\citenamefont{Russotto, Gannon,
  Kupny, Lasko, Acosta, Adamczyk, Al-Ajlan, Al-Garawi, Al-Homaidhi, Amorini
  et~al.}}]{PhysRevC_94_034608}
\bibinfo{author}{\bibfnamefont{P.}~\bibnamefont{Russotto}},
  \bibinfo{author}{\bibfnamefont{S.}~\bibnamefont{Gannon}},
  \bibinfo{author}{\bibfnamefont{S.}~\bibnamefont{Kupny}},
  \bibinfo{author}{\bibfnamefont{P.}~\bibnamefont{Lasko}},
  \bibinfo{author}{\bibfnamefont{L.}~\bibnamefont{Acosta}},
  \bibinfo{author}{\bibfnamefont{M.}~\bibnamefont{Adamczyk}},
  \bibinfo{author}{\bibfnamefont{A.}~\bibnamefont{Al-Ajlan}},
  \bibinfo{author}{\bibfnamefont{M.}~\bibnamefont{Al-Garawi}},
  \bibinfo{author}{\bibfnamefont{S.}~\bibnamefont{Al-Homaidhi}},
  \bibinfo{author}{\bibfnamefont{F.}~\bibnamefont{Amorini}},
  \bibnamefont{et~al.}, \bibinfo{journal}{Phys. Rev. C}
  \textbf{\bibinfo{volume}{94}}, \bibinfo{pages}{034608}
  (\bibinfo{year}{2016}),
  \urlprefix\url{https://link.aps.org/doi/10.1103/PhysRevC.94.034608}.

\bibitem[{\citenamefont{Estee et~al.}(2021)\citenamefont{Estee, Lynch, Tsang,
  Barney, Jhang, Tsang, Wang, Kaneko, Lee, Isobe
  et~al.}}]{PhysRevLett_126_162701}
\bibinfo{author}{\bibfnamefont{J.}~\bibnamefont{Estee}},
  \bibinfo{author}{\bibfnamefont{W.~G.} \bibnamefont{Lynch}},
  \bibinfo{author}{\bibfnamefont{C.~Y.} \bibnamefont{Tsang}},
  \bibinfo{author}{\bibfnamefont{J.}~\bibnamefont{Barney}},
  \bibinfo{author}{\bibfnamefont{G.}~\bibnamefont{Jhang}},
  \bibinfo{author}{\bibfnamefont{M.~B.} \bibnamefont{Tsang}},
  \bibinfo{author}{\bibfnamefont{R.}~\bibnamefont{Wang}},
  \bibinfo{author}{\bibfnamefont{M.}~\bibnamefont{Kaneko}},
  \bibinfo{author}{\bibfnamefont{J.~W.} \bibnamefont{Lee}},
  \bibinfo{author}{\bibfnamefont{T.}~\bibnamefont{Isobe}}, \bibnamefont{et~al.}
  (\bibinfo{collaboration}{$\mathrm{S}\ensuremath{\pi}\mathrm{RIT}$
  Collaboration}), \bibinfo{journal}{Phys. Rev. Lett.}
  \textbf{\bibinfo{volume}{126}}, \bibinfo{pages}{162701}
  (\bibinfo{year}{2021}),
  \urlprefix\url{https://link.aps.org/doi/10.1103/PhysRevLett.126.162701}.

\bibitem[{\citenamefont{Tsang et~al.}(2009)\citenamefont{Tsang, Zhang,
  Danielewicz, Famiano, Li, Lynch, and Steiner}}]{PhysRevLett_102_122701}
\bibinfo{author}{\bibfnamefont{M.~B.} \bibnamefont{Tsang}},
  \bibinfo{author}{\bibfnamefont{Y.}~\bibnamefont{Zhang}},
  \bibinfo{author}{\bibfnamefont{P.}~\bibnamefont{Danielewicz}},
  \bibinfo{author}{\bibfnamefont{M.}~\bibnamefont{Famiano}},
  \bibinfo{author}{\bibfnamefont{Z.}~\bibnamefont{Li}},
  \bibinfo{author}{\bibfnamefont{W.~G.} \bibnamefont{Lynch}}, \bibnamefont{and}
  \bibinfo{author}{\bibfnamefont{A.~W.} \bibnamefont{Steiner}},
  \bibinfo{journal}{Phys. Rev. Lett.} \textbf{\bibinfo{volume}{102}},
  \bibinfo{pages}{122701} (\bibinfo{year}{2009}),
  \urlprefix\url{https://link.aps.org/doi/10.1103/PhysRevLett.102.122701}.

\bibitem[{\citenamefont{Rami et~al.}(2000)\citenamefont{Rami, Leifels,
  de~Schauenburg, Gobbi, Hong, Alard, Andronic, Averbeck, Barret, Basrak
  et~al.}}]{Rami_PhysRevLett_84_1120}
\bibinfo{author}{\bibfnamefont{F.}~\bibnamefont{Rami}},
  \bibinfo{author}{\bibfnamefont{Y.}~\bibnamefont{Leifels}},
  \bibinfo{author}{\bibfnamefont{B.}~\bibnamefont{de~Schauenburg}},
  \bibinfo{author}{\bibfnamefont{A.}~\bibnamefont{Gobbi}},
  \bibinfo{author}{\bibfnamefont{B.}~\bibnamefont{Hong}},
  \bibinfo{author}{\bibfnamefont{J.~P.} \bibnamefont{Alard}},
  \bibinfo{author}{\bibfnamefont{A.}~\bibnamefont{Andronic}},
  \bibinfo{author}{\bibfnamefont{R.}~\bibnamefont{Averbeck}},
  \bibinfo{author}{\bibfnamefont{V.}~\bibnamefont{Barret}},
  \bibinfo{author}{\bibfnamefont{Z.}~\bibnamefont{Basrak}},
  \bibnamefont{et~al.}, \bibinfo{journal}{Phys. Rev. Lett.}
  \textbf{\bibinfo{volume}{84}}, \bibinfo{pages}{1120} (\bibinfo{year}{2000}).

\bibitem[{\citenamefont{Camaiani et~al.}(2020)\citenamefont{Camaiani,
  Piantelli, Ono, Casini, Borderie, Bougault, Ciampi, Due\~nas, Frosin,
  Frankland et~al.}}]{CAMAIANI_PRC_102_044607}
\bibinfo{author}{\bibfnamefont{A.}~\bibnamefont{Camaiani}},
  \bibinfo{author}{\bibfnamefont{S.}~\bibnamefont{Piantelli}},
  \bibinfo{author}{\bibfnamefont{A.}~\bibnamefont{Ono}},
  \bibinfo{author}{\bibfnamefont{G.}~\bibnamefont{Casini}},
  \bibinfo{author}{\bibfnamefont{B.}~\bibnamefont{Borderie}},
  \bibinfo{author}{\bibfnamefont{R.}~\bibnamefont{Bougault}},
  \bibinfo{author}{\bibfnamefont{C.}~\bibnamefont{Ciampi}},
  \bibinfo{author}{\bibfnamefont{J.~A.} \bibnamefont{Due\~nas}},
  \bibinfo{author}{\bibfnamefont{C.}~\bibnamefont{Frosin}},
  \bibinfo{author}{\bibfnamefont{J.~D.} \bibnamefont{Frankland}},
  \bibnamefont{et~al.}, \bibinfo{journal}{Phys. Rev. C}
  \textbf{\bibinfo{volume}{102}}, \bibinfo{pages}{044607}
  (\bibinfo{year}{2020}),
  \urlprefix\url{https://link.aps.org/doi/10.1103/PhysRevC.102.044607}.

\bibitem[{\citenamefont{Camaiani et~al.}(2021)\citenamefont{Camaiani, Casini,
  Piantelli, Ono, Bonnet, Alba, Barlini, Borderie, Bougault, Ciampi
  et~al.}}]{CAMAIANI_PhysRevC_103_014605}
\bibinfo{author}{\bibfnamefont{A.}~\bibnamefont{Camaiani}},
  \bibinfo{author}{\bibfnamefont{G.}~\bibnamefont{Casini}},
  \bibinfo{author}{\bibfnamefont{S.}~\bibnamefont{Piantelli}},
  \bibinfo{author}{\bibfnamefont{A.}~\bibnamefont{Ono}},
  \bibinfo{author}{\bibfnamefont{E.}~\bibnamefont{Bonnet}},
  \bibinfo{author}{\bibfnamefont{R.}~\bibnamefont{Alba}},
  \bibinfo{author}{\bibfnamefont{S.}~\bibnamefont{Barlini}},
  \bibinfo{author}{\bibfnamefont{B.}~\bibnamefont{Borderie}},
  \bibinfo{author}{\bibfnamefont{R.}~\bibnamefont{Bougault}},
  \bibinfo{author}{\bibfnamefont{C.}~\bibnamefont{Ciampi}},
  \bibnamefont{et~al.}, \bibinfo{journal}{Phys. Rev. C}
  \textbf{\bibinfo{volume}{103}}, \bibinfo{pages}{014605}
  (\bibinfo{year}{2021}),
  \urlprefix\url{https://link.aps.org/doi/10.1103/PhysRevC.103.014605}.

\bibitem[{\citenamefont{Piantelli et~al.}(2021)\citenamefont{Piantelli, Casini,
  Ono, Poggi, Pastore, Barlini, Bini, Boiano, Bonnet, Borderie
  et~al.}}]{PhysRevC_103_014603}
\bibinfo{author}{\bibfnamefont{S.}~\bibnamefont{Piantelli}},
  \bibinfo{author}{\bibfnamefont{G.}~\bibnamefont{Casini}},
  \bibinfo{author}{\bibfnamefont{A.}~\bibnamefont{Ono}},
  \bibinfo{author}{\bibfnamefont{G.}~\bibnamefont{Poggi}},
  \bibinfo{author}{\bibfnamefont{G.}~\bibnamefont{Pastore}},
  \bibinfo{author}{\bibfnamefont{S.}~\bibnamefont{Barlini}},
  \bibinfo{author}{\bibfnamefont{M.}~\bibnamefont{Bini}},
  \bibinfo{author}{\bibfnamefont{A.}~\bibnamefont{Boiano}},
  \bibinfo{author}{\bibfnamefont{E.}~\bibnamefont{Bonnet}},
  \bibinfo{author}{\bibfnamefont{B.}~\bibnamefont{Borderie}},
  \bibnamefont{et~al.} (\bibinfo{collaboration}{FAZIA Collaboration}),
  \bibinfo{journal}{Phys. Rev. C} \textbf{\bibinfo{volume}{103}},
  \bibinfo{pages}{014603} (\bibinfo{year}{2021}),
  \urlprefix\url{https://link.aps.org/doi/10.1103/PhysRevC.103.014603}.

\bibitem[{\citenamefont{Ciampi et~al.}(2022)\citenamefont{Ciampi, Piantelli,
  Casini, Pasquali, Quicray, Baldesi, Barlini, Borderie, Bougault, Camaiani
  et~al.}}]{Ciampi_PRC_106_024603}
\bibinfo{author}{\bibfnamefont{C.}~\bibnamefont{Ciampi}},
  \bibinfo{author}{\bibfnamefont{S.}~\bibnamefont{Piantelli}},
  \bibinfo{author}{\bibfnamefont{G.}~\bibnamefont{Casini}},
  \bibinfo{author}{\bibfnamefont{G.}~\bibnamefont{Pasquali}},
  \bibinfo{author}{\bibfnamefont{J.}~\bibnamefont{Quicray}},
  \bibinfo{author}{\bibfnamefont{L.}~\bibnamefont{Baldesi}},
  \bibinfo{author}{\bibfnamefont{S.}~\bibnamefont{Barlini}},
  \bibinfo{author}{\bibfnamefont{B.}~\bibnamefont{Borderie}},
  \bibinfo{author}{\bibfnamefont{R.}~\bibnamefont{Bougault}},
  \bibinfo{author}{\bibfnamefont{A.}~\bibnamefont{Camaiani}},
  \bibnamefont{et~al.}, \bibinfo{journal}{Phys. Rev. C}
  \textbf{\bibinfo{volume}{106}}, \bibinfo{pages}{024603}
  (\bibinfo{year}{2022}).

\bibitem[{\citenamefont{Fable et~al.}(2023)\citenamefont{Fable, Chbihi,
  Frankland, Napolitani, Verde, Bonnet, Borderie, Bougault, Galichet, G\'enard
  et~al.}}]{Fable_PRC_107_014604}
\bibinfo{author}{\bibfnamefont{Q.}~\bibnamefont{Fable}},
  \bibinfo{author}{\bibfnamefont{A.}~\bibnamefont{Chbihi}},
  \bibinfo{author}{\bibfnamefont{J.~D.} \bibnamefont{Frankland}},
  \bibinfo{author}{\bibfnamefont{P.}~\bibnamefont{Napolitani}},
  \bibinfo{author}{\bibfnamefont{G.}~\bibnamefont{Verde}},
  \bibinfo{author}{\bibfnamefont{E.}~\bibnamefont{Bonnet}},
  \bibinfo{author}{\bibfnamefont{B.}~\bibnamefont{Borderie}},
  \bibinfo{author}{\bibfnamefont{R.}~\bibnamefont{Bougault}},
  \bibinfo{author}{\bibfnamefont{E.}~\bibnamefont{Galichet}},
  \bibinfo{author}{\bibfnamefont{T.}~\bibnamefont{G\'enard}},
  \bibnamefont{et~al.} (\bibinfo{collaboration}{INDRA Collaboration}),
  \bibinfo{journal}{Phys. Rev. C} \textbf{\bibinfo{volume}{107}},
  \bibinfo{pages}{014604} (\bibinfo{year}{2023}),
  \urlprefix\url{https://link.aps.org/doi/10.1103/PhysRevC.107.014604}.

\bibitem[{\citenamefont{Pouthas et~al.}(1995)\citenamefont{Pouthas, Borderie,
  Dayras, Plagnol, Rivet, Saint-Laurent, Steckmeyer, Auger, Bacri, Barbey
  et~al.}}]{I3-Pou95}
\bibinfo{author}{\bibfnamefont{J.}~\bibnamefont{Pouthas}},
  \bibinfo{author}{\bibfnamefont{B.}~\bibnamefont{Borderie}},
  \bibinfo{author}{\bibfnamefont{R.}~\bibnamefont{Dayras}},
  \bibinfo{author}{\bibfnamefont{E.}~\bibnamefont{Plagnol}},
  \bibinfo{author}{\bibfnamefont{M.~F.} \bibnamefont{Rivet}},
  \bibinfo{author}{\bibfnamefont{F.}~\bibnamefont{Saint-Laurent}},
  \bibinfo{author}{\bibfnamefont{J.~C.} \bibnamefont{Steckmeyer}},
  \bibinfo{author}{\bibfnamefont{G.}~\bibnamefont{Auger}},
  \bibinfo{author}{\bibfnamefont{C.~O.} \bibnamefont{Bacri}},
  \bibinfo{author}{\bibfnamefont{S.}~\bibnamefont{Barbey}},
  \bibnamefont{et~al.}, \bibinfo{journal}{Nucl. Instr. and Meth. in Phys. Res.}
  \textbf{\bibinfo{volume}{A 357}}, \bibinfo{pages}{418}
  (\bibinfo{year}{1995}).

\bibitem[{\citenamefont{Pouthas et~al.}(1996)\citenamefont{Pouthas, Bertaut,
  Borderie, Bourgault, Cahan, Carles, Charlet, Cussol, Dayras, Engrand
  et~al.}}]{I5-Pou96}
\bibinfo{author}{\bibfnamefont{J.}~\bibnamefont{Pouthas}},
  \bibinfo{author}{\bibfnamefont{A.}~\bibnamefont{Bertaut}},
  \bibinfo{author}{\bibfnamefont{B.}~\bibnamefont{Borderie}},
  \bibinfo{author}{\bibfnamefont{P.}~\bibnamefont{Bourgault}},
  \bibinfo{author}{\bibfnamefont{B.}~\bibnamefont{Cahan}},
  \bibinfo{author}{\bibfnamefont{G.}~\bibnamefont{Carles}},
  \bibinfo{author}{\bibfnamefont{D.}~\bibnamefont{Charlet}},
  \bibinfo{author}{\bibfnamefont{D.}~\bibnamefont{Cussol}},
  \bibinfo{author}{\bibfnamefont{R.}~\bibnamefont{Dayras}},
  \bibinfo{author}{\bibfnamefont{M.}~\bibnamefont{Engrand}},
  \bibnamefont{et~al.}, \bibinfo{journal}{Nucl. Instr. and Meth. in Phys. Res.}
  \textbf{\bibinfo{volume}{A 369}}, \bibinfo{pages}{222}
  (\bibinfo{year}{1996}).

\bibitem[{\citenamefont{Pullanhiotan et~al.}(2008)\citenamefont{Pullanhiotan,
  Rejmund, Navin, Mittig, and Bhattacharyya}}]{PULLANHIOTAN2008343}
\bibinfo{author}{\bibfnamefont{S.}~\bibnamefont{Pullanhiotan}},
  \bibinfo{author}{\bibfnamefont{M.}~\bibnamefont{Rejmund}},
  \bibinfo{author}{\bibfnamefont{A.}~\bibnamefont{Navin}},
  \bibinfo{author}{\bibfnamefont{W.}~\bibnamefont{Mittig}}, \bibnamefont{and}
  \bibinfo{author}{\bibfnamefont{S.}~\bibnamefont{Bhattacharyya}},
  \bibinfo{journal}{Nuclear Instruments and Methods in Physics Research Section
  A: Accelerators, Spectrometers, Detectors and Associated Equipment}
  \textbf{\bibinfo{volume}{593}}, \bibinfo{pages}{343} (\bibinfo{year}{2008}),
  ISSN \bibinfo{issn}{0168-9002},
  \urlprefix\url{https://www.sciencedirect.com/science/article/pii/S0168900208007080}.

\bibitem[{\citenamefont{Lemasson and Rejmund}(2023)}]{LEMASSON2023168407}
\bibinfo{author}{\bibfnamefont{A.}~\bibnamefont{Lemasson}} \bibnamefont{and}
  \bibinfo{author}{\bibfnamefont{M.}~\bibnamefont{Rejmund}},
  \bibinfo{journal}{Nuclear Instruments and Methods in Physics Research Section
  A: Accelerators, Spectrometers, Detectors and Associated Equipment}
  \textbf{\bibinfo{volume}{1054}}, \bibinfo{pages}{168407}
  (\bibinfo{year}{2023}), ISSN \bibinfo{issn}{0168-9002},
  \urlprefix\url{https://www.sciencedirect.com/science/article/pii/S0168900223003972}.

\bibitem[{\citenamefont{Fable et~al.}(2022)\citenamefont{Fable, Chbihi,
  Boisjoli, Frankland, Le~F\`evre, Le~Neindre, Marini, Verde, Ademard, Bardelli
  et~al.}}]{Fable_PRC_106_024605}
\bibinfo{author}{\bibfnamefont{Q.}~\bibnamefont{Fable}},
  \bibinfo{author}{\bibfnamefont{A.}~\bibnamefont{Chbihi}},
  \bibinfo{author}{\bibfnamefont{M.}~\bibnamefont{Boisjoli}},
  \bibinfo{author}{\bibfnamefont{J.~D.} \bibnamefont{Frankland}},
  \bibinfo{author}{\bibfnamefont{A.}~\bibnamefont{Le~F\`evre}},
  \bibinfo{author}{\bibfnamefont{N.}~\bibnamefont{Le~Neindre}},
  \bibinfo{author}{\bibfnamefont{P.}~\bibnamefont{Marini}},
  \bibinfo{author}{\bibfnamefont{G.}~\bibnamefont{Verde}},
  \bibinfo{author}{\bibfnamefont{G.}~\bibnamefont{Ademard}},
  \bibinfo{author}{\bibfnamefont{L.}~\bibnamefont{Bardelli}},
  \bibnamefont{et~al.}, \bibinfo{journal}{Phys. Rev. C}
  \textbf{\bibinfo{volume}{106}}, \bibinfo{pages}{024605}
  (\bibinfo{year}{2022}).

\bibitem[{\citenamefont{Ono et~al.}(1992)\citenamefont{Ono, Horiuchi, Maruyama,
  and Ohnishi}}]{ONO_PTP_87_5_1185}
\bibinfo{author}{\bibfnamefont{A.}~\bibnamefont{Ono}},
  \bibinfo{author}{\bibfnamefont{H.}~\bibnamefont{Horiuchi}},
  \bibinfo{author}{\bibfnamefont{T.}~\bibnamefont{Maruyama}}, \bibnamefont{and}
  \bibinfo{author}{\bibfnamefont{A.}~\bibnamefont{Ohnishi}},
  \bibinfo{journal}{Progress of Theoretical Physics}
  \textbf{\bibinfo{volume}{87}}, \bibinfo{pages}{1185} (\bibinfo{year}{1992}),
  ISSN \bibinfo{issn}{0033-068X},
  \eprint{https://academic.oup.com/ptp/article-pdf/87/5/1185/5272175/87-5-1185.pdf},
  \urlprefix\url{https://doi.org/10.1143/ptp/87.5.1185}.

\bibitem[{\citenamefont{Kanada-En'yo et~al.}(2003)\citenamefont{Kanada-En'yo,
  Kimura, and Horiuchi}}]{KANADAENYO2003497}
\bibinfo{author}{\bibfnamefont{Y.}~\bibnamefont{Kanada-En'yo}},
  \bibinfo{author}{\bibfnamefont{M.}~\bibnamefont{Kimura}}, \bibnamefont{and}
  \bibinfo{author}{\bibfnamefont{H.}~\bibnamefont{Horiuchi}},
  \bibinfo{journal}{Comptes Rendus Physique} \textbf{\bibinfo{volume}{4}},
  \bibinfo{pages}{497} (\bibinfo{year}{2003}), ISSN \bibinfo{issn}{1631-0705},
  \urlprefix\url{https://www.sciencedirect.com/science/article/pii/S1631070503000628}.

\bibitem[{\citenamefont{Ono}(2013)}]{Ono_2013}
\bibinfo{author}{\bibfnamefont{A.}~\bibnamefont{Ono}},
  \bibinfo{journal}{Journal of Physics: Conference Series}
  \textbf{\bibinfo{volume}{420}}, \bibinfo{pages}{012103}
  (\bibinfo{year}{2013}),
  \urlprefix\url{https://dx.doi.org/10.1088/1742-6596/420/1/012103}.

\bibitem[{\citenamefont{Wolter et~al.}(2022)\citenamefont{Wolter, Colonna,
  Cozma, Danielewicz, Ko, Kumar, Ono, Tsang, Xu, Zhang
  et~al.}}]{WOLTER2022103962}
\bibinfo{author}{\bibfnamefont{H.}~\bibnamefont{Wolter}},
  \bibinfo{author}{\bibfnamefont{M.}~\bibnamefont{Colonna}},
  \bibinfo{author}{\bibfnamefont{D.}~\bibnamefont{Cozma}},
  \bibinfo{author}{\bibfnamefont{P.}~\bibnamefont{Danielewicz}},
  \bibinfo{author}{\bibfnamefont{C.~M.} \bibnamefont{Ko}},
  \bibinfo{author}{\bibfnamefont{R.}~\bibnamefont{Kumar}},
  \bibinfo{author}{\bibfnamefont{A.}~\bibnamefont{Ono}},
  \bibinfo{author}{\bibfnamefont{M.~B.} \bibnamefont{Tsang}},
  \bibinfo{author}{\bibfnamefont{J.}~\bibnamefont{Xu}},
  \bibinfo{author}{\bibfnamefont{Y.-X.} \bibnamefont{Zhang}},
  \bibnamefont{et~al.}, \bibinfo{journal}{Progress in Particle and Nuclear
  Physics} \textbf{\bibinfo{volume}{125}}, \bibinfo{pages}{103962}
  (\bibinfo{year}{2022}), ISSN \bibinfo{issn}{0146-6410},
  \urlprefix\url{https://www.sciencedirect.com/science/article/pii/S0146641022000230}.

\bibitem[{\citenamefont{Frosin et~al.}(2023)\citenamefont{Frosin, Piantelli,
  Casini, Ono, Camaiani, Baldesi, Barlini, Borderie, Bougault, Ciampi
  et~al.}}]{PhysRevC_107_044614}
\bibinfo{author}{\bibfnamefont{C.}~\bibnamefont{Frosin}},
  \bibinfo{author}{\bibfnamefont{S.}~\bibnamefont{Piantelli}},
  \bibinfo{author}{\bibfnamefont{G.}~\bibnamefont{Casini}},
  \bibinfo{author}{\bibfnamefont{A.}~\bibnamefont{Ono}},
  \bibinfo{author}{\bibfnamefont{A.}~\bibnamefont{Camaiani}},
  \bibinfo{author}{\bibfnamefont{L.}~\bibnamefont{Baldesi}},
  \bibinfo{author}{\bibfnamefont{S.}~\bibnamefont{Barlini}},
  \bibinfo{author}{\bibfnamefont{B.}~\bibnamefont{Borderie}},
  \bibinfo{author}{\bibfnamefont{R.}~\bibnamefont{Bougault}},
  \bibinfo{author}{\bibfnamefont{C.}~\bibnamefont{Ciampi}},
  \bibnamefont{et~al.} (\bibinfo{collaboration}{INDRA-FAZIA Collaboration}),
  \bibinfo{journal}{Phys. Rev. C} \textbf{\bibinfo{volume}{107}},
  \bibinfo{pages}{044614} (\bibinfo{year}{2023}),
  \urlprefix\url{https://link.aps.org/doi/10.1103/PhysRevC.107.044614}.

\bibitem[{\citenamefont{Piantelli et~al.}(2017)\citenamefont{Piantelli,
  Valdr\'e, Barlini, Casini, Colonna, Baiocco, Bini, Bruno, Camaiani, Carboni
  et~al.}}]{PhysRevC_96_034622}
\bibinfo{author}{\bibfnamefont{S.}~\bibnamefont{Piantelli}},
  \bibinfo{author}{\bibfnamefont{S.}~\bibnamefont{Valdr\'e}},
  \bibinfo{author}{\bibfnamefont{S.}~\bibnamefont{Barlini}},
  \bibinfo{author}{\bibfnamefont{G.}~\bibnamefont{Casini}},
  \bibinfo{author}{\bibfnamefont{M.}~\bibnamefont{Colonna}},
  \bibinfo{author}{\bibfnamefont{G.}~\bibnamefont{Baiocco}},
  \bibinfo{author}{\bibfnamefont{M.}~\bibnamefont{Bini}},
  \bibinfo{author}{\bibfnamefont{M.}~\bibnamefont{Bruno}},
  \bibinfo{author}{\bibfnamefont{A.}~\bibnamefont{Camaiani}},
  \bibinfo{author}{\bibfnamefont{S.}~\bibnamefont{Carboni}},
  \bibnamefont{et~al.}, \bibinfo{journal}{Phys. Rev. C}
  \textbf{\bibinfo{volume}{96}}, \bibinfo{pages}{034622}
  (\bibinfo{year}{2017}),
  \urlprefix\url{https://link.aps.org/doi/10.1103/PhysRevC.96.034622}.

\bibitem[{\citenamefont{Ono}(2019)}]{ONO2019139}
\bibinfo{author}{\bibfnamefont{A.}~\bibnamefont{Ono}},
  \bibinfo{journal}{Progress in Particle and Nuclear Physics}
  \textbf{\bibinfo{volume}{105}}, \bibinfo{pages}{139} (\bibinfo{year}{2019}),
  ISSN \bibinfo{issn}{0146-6410},
  \urlprefix\url{https://www.sciencedirect.com/science/article/pii/S0146641018300863}.

\bibitem[{\citenamefont{Ono}(1999)}]{PhysRevC_59_853}
\bibinfo{author}{\bibfnamefont{A.}~\bibnamefont{Ono}}, \bibinfo{journal}{Phys.
  Rev. C} \textbf{\bibinfo{volume}{59}}, \bibinfo{pages}{853}
  (\bibinfo{year}{1999}),
  \urlprefix\url{https://link.aps.org/doi/10.1103/PhysRevC.59.853}.

\bibitem[{\citenamefont{Wada et~al.}(1998)\citenamefont{Wada, Hagel, Cibor, Li,
  Marie, Shen, Zhao, Natowitz, and Ono}}]{WADA19986}
\bibinfo{author}{\bibfnamefont{R.}~\bibnamefont{Wada}},
  \bibinfo{author}{\bibfnamefont{K.}~\bibnamefont{Hagel}},
  \bibinfo{author}{\bibfnamefont{J.}~\bibnamefont{Cibor}},
  \bibinfo{author}{\bibfnamefont{J.}~\bibnamefont{Li}},
  \bibinfo{author}{\bibfnamefont{N.}~\bibnamefont{Marie}},
  \bibinfo{author}{\bibfnamefont{W.}~\bibnamefont{Shen}},
  \bibinfo{author}{\bibfnamefont{Y.}~\bibnamefont{Zhao}},
  \bibinfo{author}{\bibfnamefont{J.}~\bibnamefont{Natowitz}}, \bibnamefont{and}
  \bibinfo{author}{\bibfnamefont{A.}~\bibnamefont{Ono}},
  \bibinfo{journal}{Physics Letters B} \textbf{\bibinfo{volume}{422}},
  \bibinfo{pages}{6} (\bibinfo{year}{1998}), ISSN \bibinfo{issn}{0370-2693},
  \urlprefix\url{https://www.sciencedirect.com/science/article/pii/S0370269398000331}.

\bibitem[{\citenamefont{Ono et~al.}(2004)\citenamefont{Ono, Danielewicz,
  Friedman, Lynch, and Tsang}}]{PhysRevC_70_041604}
\bibinfo{author}{\bibfnamefont{A.}~\bibnamefont{Ono}},
  \bibinfo{author}{\bibfnamefont{P.}~\bibnamefont{Danielewicz}},
  \bibinfo{author}{\bibfnamefont{W.~A.} \bibnamefont{Friedman}},
  \bibinfo{author}{\bibfnamefont{W.~G.} \bibnamefont{Lynch}}, \bibnamefont{and}
  \bibinfo{author}{\bibfnamefont{M.~B.} \bibnamefont{Tsang}},
  \bibinfo{journal}{Phys. Rev. C} \textbf{\bibinfo{volume}{70}},
  \bibinfo{pages}{041604} (\bibinfo{year}{2004}),
  \urlprefix\url{https://link.aps.org/doi/10.1103/PhysRevC.70.041604}.

\bibitem[{\citenamefont{Decharg\'e and Gogny}(1980)}]{PhysRevC_21_1568}
\bibinfo{author}{\bibfnamefont{J.}~\bibnamefont{Decharg\'e}} \bibnamefont{and}
  \bibinfo{author}{\bibfnamefont{D.}~\bibnamefont{Gogny}},
  \bibinfo{journal}{Phys. Rev. C} \textbf{\bibinfo{volume}{21}},
  \bibinfo{pages}{1568} (\bibinfo{year}{1980}),
  \urlprefix\url{https://link.aps.org/doi/10.1103/PhysRevC.21.1568}.

\bibitem[{\citenamefont{Chabanat et~al.}(1997)\citenamefont{Chabanat, Bonche,
  Haensel, Meyer, and Schaeffer}}]{CHABANAT1997710}
\bibinfo{author}{\bibfnamefont{E.}~\bibnamefont{Chabanat}},
  \bibinfo{author}{\bibfnamefont{P.}~\bibnamefont{Bonche}},
  \bibinfo{author}{\bibfnamefont{P.}~\bibnamefont{Haensel}},
  \bibinfo{author}{\bibfnamefont{J.}~\bibnamefont{Meyer}}, \bibnamefont{and}
  \bibinfo{author}{\bibfnamefont{R.}~\bibnamefont{Schaeffer}},
  \bibinfo{journal}{Nuclear Physics A} \textbf{\bibinfo{volume}{627}},
  \bibinfo{pages}{710} (\bibinfo{year}{1997}), ISSN \bibinfo{issn}{0375-9474},
  \urlprefix\url{https://www.sciencedirect.com/science/article/pii/S0375947497005964}.

\bibitem[{\citenamefont{Lopez et~al.}(2014)\citenamefont{Lopez, Durand, Lehaut,
  Borderie, Frankland, Rivet, Bougault, Chbihi, Galichet, Guinet
  et~al.}}]{PhysRevC_90_064602}
\bibinfo{author}{\bibfnamefont{O.}~\bibnamefont{Lopez}},
  \bibinfo{author}{\bibfnamefont{D.}~\bibnamefont{Durand}},
  \bibinfo{author}{\bibfnamefont{G.}~\bibnamefont{Lehaut}},
  \bibinfo{author}{\bibfnamefont{B.}~\bibnamefont{Borderie}},
  \bibinfo{author}{\bibfnamefont{J.~D.} \bibnamefont{Frankland}},
  \bibinfo{author}{\bibfnamefont{M.~F.} \bibnamefont{Rivet}},
  \bibinfo{author}{\bibfnamefont{R.}~\bibnamefont{Bougault}},
  \bibinfo{author}{\bibfnamefont{A.}~\bibnamefont{Chbihi}},
  \bibinfo{author}{\bibfnamefont{E.}~\bibnamefont{Galichet}},
  \bibinfo{author}{\bibfnamefont{D.}~\bibnamefont{Guinet}},
  \bibnamefont{et~al.} (\bibinfo{collaboration}{INDRA Collaboration}),
  \bibinfo{journal}{Phys. Rev. C} \textbf{\bibinfo{volume}{90}},
  \bibinfo{pages}{064602} (\bibinfo{year}{2014}),
  \urlprefix\url{https://link.aps.org/doi/10.1103/PhysRevC.90.064602}.

\bibitem[{\citenamefont{Charity}(2010)}]{PhysRevC_82_014610}
\bibinfo{author}{\bibfnamefont{R.~J.} \bibnamefont{Charity}},
  \bibinfo{journal}{Phys. Rev. C} \textbf{\bibinfo{volume}{82}},
  \bibinfo{pages}{014610} (\bibinfo{year}{2010}),
  \urlprefix\url{https://link.aps.org/doi/10.1103/PhysRevC.82.014610}.

\bibitem[{\citenamefont{Piantelli et~al.}(2019)\citenamefont{Piantelli, Olmi,
  Maurenzig, Ono, Bini, Casini, Pasquali, Mangiarotti, Poggi, Stefanini
  et~al.}}]{PhysRevC_99_064616}
\bibinfo{author}{\bibfnamefont{S.}~\bibnamefont{Piantelli}},
  \bibinfo{author}{\bibfnamefont{A.}~\bibnamefont{Olmi}},
  \bibinfo{author}{\bibfnamefont{P.~R.} \bibnamefont{Maurenzig}},
  \bibinfo{author}{\bibfnamefont{A.}~\bibnamefont{Ono}},
  \bibinfo{author}{\bibfnamefont{M.}~\bibnamefont{Bini}},
  \bibinfo{author}{\bibfnamefont{G.}~\bibnamefont{Casini}},
  \bibinfo{author}{\bibfnamefont{G.}~\bibnamefont{Pasquali}},
  \bibinfo{author}{\bibfnamefont{A.}~\bibnamefont{Mangiarotti}},
  \bibinfo{author}{\bibfnamefont{G.}~\bibnamefont{Poggi}},
  \bibinfo{author}{\bibfnamefont{A.~A.} \bibnamefont{Stefanini}},
  \bibnamefont{et~al.}, \bibinfo{journal}{Phys. Rev. C}
  \textbf{\bibinfo{volume}{99}}, \bibinfo{pages}{064616}
  (\bibinfo{year}{2019}),
  \urlprefix\url{https://link.aps.org/doi/10.1103/PhysRevC.99.064616}.

\bibitem[{\citenamefont{Frankland et~al.}(2023)\citenamefont{Frankland, Bonnet,
  and Gruyer}}]{KaliVeda}
\bibinfo{author}{\bibfnamefont{J.}~\bibnamefont{Frankland}},
  \bibinfo{author}{\bibfnamefont{E.}~\bibnamefont{Bonnet}}, \bibnamefont{and}
  \bibinfo{author}{\bibfnamefont{D.}~\bibnamefont{Gruyer}},
  \emph{\bibinfo{title}{Kaliveda: Heavy-ion collisions analysis toolkit}}
  (\bibinfo{year}{2023}),
  \urlprefix\url{https://doi.org/10.5281/zenodo.7664901}.

\bibitem[{\citenamefont{Piantelli et~al.}(2013)\citenamefont{Piantelli, Casini,
  Maurenzig, Olmi, Barlini, Bini, Carboni, Pasquali, Poggi, Stefanini
  et~al.}}]{PhysRevC_88_064607}
\bibinfo{author}{\bibfnamefont{S.}~\bibnamefont{Piantelli}},
  \bibinfo{author}{\bibfnamefont{G.}~\bibnamefont{Casini}},
  \bibinfo{author}{\bibfnamefont{P.~R.} \bibnamefont{Maurenzig}},
  \bibinfo{author}{\bibfnamefont{A.}~\bibnamefont{Olmi}},
  \bibinfo{author}{\bibfnamefont{S.}~\bibnamefont{Barlini}},
  \bibinfo{author}{\bibfnamefont{M.}~\bibnamefont{Bini}},
  \bibinfo{author}{\bibfnamefont{S.}~\bibnamefont{Carboni}},
  \bibinfo{author}{\bibfnamefont{G.}~\bibnamefont{Pasquali}},
  \bibinfo{author}{\bibfnamefont{G.}~\bibnamefont{Poggi}},
  \bibinfo{author}{\bibfnamefont{A.~A.} \bibnamefont{Stefanini}},
  \bibnamefont{et~al.} (\bibinfo{collaboration}{FAZIA Collaboration}),
  \bibinfo{journal}{Phys. Rev. C} \textbf{\bibinfo{volume}{88}},
  \bibinfo{pages}{064607} (\bibinfo{year}{2013}),
  \urlprefix\url{https://link.aps.org/doi/10.1103/PhysRevC.88.064607}.

\bibitem[{\citenamefont{Charity}(1998)}]{PhysRevC_58_1073}
\bibinfo{author}{\bibfnamefont{R.~J.} \bibnamefont{Charity}},
  \bibinfo{journal}{Phys. Rev. C} \textbf{\bibinfo{volume}{58}},
  \bibinfo{pages}{1073} (\bibinfo{year}{1998}),
  \urlprefix\url{https://link.aps.org/doi/10.1103/PhysRevC.58.1073}.

\bibitem[{\citenamefont{Charity and Sobotka}(2005)}]{PhysRevC_71_024310}
\bibinfo{author}{\bibfnamefont{R.~J.} \bibnamefont{Charity}} \bibnamefont{and}
  \bibinfo{author}{\bibfnamefont{L.~G.} \bibnamefont{Sobotka}},
  \bibinfo{journal}{Phys. Rev. C} \textbf{\bibinfo{volume}{71}},
  \bibinfo{pages}{024310} (\bibinfo{year}{2005}),
  \urlprefix\url{https://link.aps.org/doi/10.1103/PhysRevC.71.024310}.

\bibitem[{\citenamefont{Das et~al.}(2018)\citenamefont{Das, Giacalone, Monard,
  and Ollitrault}}]{PhysRevC_97_014905}
\bibinfo{author}{\bibfnamefont{S.~J.} \bibnamefont{Das}},
  \bibinfo{author}{\bibfnamefont{G.}~\bibnamefont{Giacalone}},
  \bibinfo{author}{\bibfnamefont{P.-A.} \bibnamefont{Monard}},
  \bibnamefont{and} \bibinfo{author}{\bibfnamefont{J.-Y.}
  \bibnamefont{Ollitrault}}, \bibinfo{journal}{Phys. Rev. C}
  \textbf{\bibinfo{volume}{97}}, \bibinfo{pages}{014905}
  (\bibinfo{year}{2018}),
  \urlprefix\url{https://link.aps.org/doi/10.1103/PhysRevC.97.014905}.

\bibitem[{\citenamefont{Rogly et~al.}(2018)\citenamefont{Rogly, Giacalone, and
  Ollitrault}}]{PhysRevC_98_024902}
\bibinfo{author}{\bibfnamefont{R.}~\bibnamefont{Rogly}},
  \bibinfo{author}{\bibfnamefont{G.}~\bibnamefont{Giacalone}},
  \bibnamefont{and} \bibinfo{author}{\bibfnamefont{J.-Y.}
  \bibnamefont{Ollitrault}}, \bibinfo{journal}{Phys. Rev. C}
  \textbf{\bibinfo{volume}{98}}, \bibinfo{pages}{024902}
  (\bibinfo{year}{2018}),
  \urlprefix\url{https://link.aps.org/doi/10.1103/PhysRevC.98.024902}.

\bibitem[{\citenamefont{Frankland et~al.}(2021)\citenamefont{Frankland, Gruyer,
  Bonnet, Borderie, Bougault, Chbihi, Ducret, Durand, Fable, Henri
  et~al.}}]{PhysRevC_104_034609}
\bibinfo{author}{\bibfnamefont{J.~D.} \bibnamefont{Frankland}},
  \bibinfo{author}{\bibfnamefont{D.}~\bibnamefont{Gruyer}},
  \bibinfo{author}{\bibfnamefont{E.}~\bibnamefont{Bonnet}},
  \bibinfo{author}{\bibfnamefont{B.}~\bibnamefont{Borderie}},
  \bibinfo{author}{\bibfnamefont{R.}~\bibnamefont{Bougault}},
  \bibinfo{author}{\bibfnamefont{A.}~\bibnamefont{Chbihi}},
  \bibinfo{author}{\bibfnamefont{J.~E.} \bibnamefont{Ducret}},
  \bibinfo{author}{\bibfnamefont{D.}~\bibnamefont{Durand}},
  \bibinfo{author}{\bibfnamefont{Q.}~\bibnamefont{Fable}},
  \bibinfo{author}{\bibfnamefont{M.}~\bibnamefont{Henri}}, \bibnamefont{et~al.}
  (\bibinfo{collaboration}{INDRA Collaboration}), \bibinfo{journal}{Phys. Rev.
  C} \textbf{\bibinfo{volume}{104}}, \bibinfo{pages}{034609}
  (\bibinfo{year}{2021}),
  \urlprefix\url{https://link.aps.org/doi/10.1103/PhysRevC.104.034609}.

\bibitem[{\citenamefont{Ogul et~al.}(2023)\citenamefont{Ogul, Botvina,
  Bleicher, Buyukcizmeci, Ergun, Imal, Leifels, and
  Trautmann}}]{PhysRevC_107_054606}
\bibinfo{author}{\bibfnamefont{R.}~\bibnamefont{Ogul}},
  \bibinfo{author}{\bibfnamefont{A.~S.} \bibnamefont{Botvina}},
  \bibinfo{author}{\bibfnamefont{M.}~\bibnamefont{Bleicher}},
  \bibinfo{author}{\bibfnamefont{N.}~\bibnamefont{Buyukcizmeci}},
  \bibinfo{author}{\bibfnamefont{A.}~\bibnamefont{Ergun}},
  \bibinfo{author}{\bibfnamefont{H.}~\bibnamefont{Imal}},
  \bibinfo{author}{\bibfnamefont{Y.}~\bibnamefont{Leifels}}, \bibnamefont{and}
  \bibinfo{author}{\bibfnamefont{W.}~\bibnamefont{Trautmann}},
  \bibinfo{journal}{Phys. Rev. C} \textbf{\bibinfo{volume}{107}},
  \bibinfo{pages}{054606} (\bibinfo{year}{2023}),
  \urlprefix\url{https://link.aps.org/doi/10.1103/PhysRevC.107.054606}.

\bibitem[{\citenamefont{Tsang et~al.}(2004)\citenamefont{Tsang, Liu, Shi,
  Danielewicz, Gelbke, Liu, Lynch, Tan, Verde, Wagner
  et~al.}}]{Tsang2004:isospindiffusion}
\bibinfo{author}{\bibfnamefont{M.~B.} \bibnamefont{Tsang}},
  \bibinfo{author}{\bibfnamefont{T.~X.} \bibnamefont{Liu}},
  \bibinfo{author}{\bibfnamefont{L.}~\bibnamefont{Shi}},
  \bibinfo{author}{\bibfnamefont{P.}~\bibnamefont{Danielewicz}},
  \bibinfo{author}{\bibfnamefont{C.~K.} \bibnamefont{Gelbke}},
  \bibinfo{author}{\bibfnamefont{X.~D.} \bibnamefont{Liu}},
  \bibinfo{author}{\bibfnamefont{W.~G.} \bibnamefont{Lynch}},
  \bibinfo{author}{\bibfnamefont{W.~P.} \bibnamefont{Tan}},
  \bibinfo{author}{\bibfnamefont{G.}~\bibnamefont{Verde}},
  \bibinfo{author}{\bibfnamefont{A.}~\bibnamefont{Wagner}},
  \bibnamefont{et~al.}, \bibinfo{journal}{Phys. Rev. Lett.}
  \textbf{\bibinfo{volume}{92}}, \bibinfo{pages}{062701}
  (\bibinfo{year}{2004}).

\bibitem[{\citenamefont{{Colonna, Maria} et~al.}(2014)\citenamefont{{Colonna,
  Maria}, {Baran, Virgil}, and {Di Toro, Massimo}}}]{Colonna_EPJA50}
\bibinfo{author}{\bibnamefont{{Colonna, Maria}}},
  \bibinfo{author}{\bibnamefont{{Baran, Virgil}}}, \bibnamefont{and}
  \bibinfo{author}{\bibnamefont{{Di Toro, Massimo}}}, \bibinfo{journal}{Eur.
  Phys. J. A} \textbf{\bibinfo{volume}{50}}, \bibinfo{pages}{30}
  (\bibinfo{year}{2014}).

\end{thebibliography}

\end{document}